\documentclass[Afour,sageh,times]{sagej}

\usepackage{moreverb,url}

\usepackage[title]{appendix}

\usepackage{multicol}

\usepackage{balance}

\usepackage[colorlinks,bookmarksopen,bookmarksnumbered,citecolor=red,urlcolor=red]{hyperref}

\newcommand\BibTeX{{\rmfamily B\kern-.05em \textsc{i\kern-.025em b}\kern-.08em
T\kern-.1667em\lower.7ex\hbox{E}\kern-.125emX}}

\def\volumeyear{2023}

\usepackage{amsmath}
\usepackage{siunitx}
\usepackage{todonotes}
\usepackage{ulem}
\usepackage{pgfplots}
\usepackage{amssymb}
\usepackage{booktabs}

\newcommand{\FK}[1]{\textcolor{black}{ #1}}
\mathchardef\ordinarycolon\mathcode`\:
\mathchardef\ordinaryequal\mathcode`\=
\mathcode`\:=\string"8000
\begingroup \catcode`\:=\active
\gdef:{\mathrel{\mathop\ordinarycolon}}
\endgroup
\begingroup \catcode`\==\active
\gdef={\mathrel{\mathop\ordinaryequal}}
\endgroup

\newcommand{\mycolor}[1]{ {\color{black}{#1}} }
\newcommand{\foeppl}[1]{\langle #1 \rangle }
\DeclareMathOperator{\sign}{sign}
\usepackage{ifthen}
\usepackage{float}

\usepackage{tikz}
\usetikzlibrary{calc}
\usepackage{subcaption}

\usetikzlibrary{shapes,arrows, chains, quotes, positioning}

\tikzstyle{decision} = [diamond, draw, fill=blue!20, text width=6.5em, text badly centered, node distance=4cm, inner sep=0pt]
\tikzstyle{block} = [rectangle, draw, fill=blue!20, text width=5em, text centered, rounded corners, minimum height=4em]
\tikzstyle{line} = [draw, -latex    ]
\tikzstyle{cloud} = [draw, ellipse,fill=red!20, node distance=2cm, minimum height=2em]
\tikzstyle{base}= [draw, minimum width=4.5em, align=flush center, outer sep=0pt]
\tikzstyle{empty}=[text width=26em]

\def\orcid#1{\kern .08em\href{https://orcid.org/#1}{\includegraphics[keepaspectratio,width=0.7em]{orcid.png}}}

\begin{document}

\runninghead{Saadi, Kölzow, Kontermann, Oechsner and Gottschalk}

\title{Uncertainty quantification for damage mechanics models using the bootstrap method}

\author{Mohamed Saadi\affilnum{1}, Felix Kölzow\affilnum{2}, Christian Kontermann\affilnum{2}, Matthias Oechsner\affilnum{2} and Hanno Gottschalk\affilnum{1}}

\affiliation{
\affilnum{1} Faculty II - Mathematics and Natural Sciences, Technical University of Berlin, Straße des 17. Juni 135, 10623 Berlin, Germany\\
\affilnum{2} Department and Institute of Materials Science, Technical University of Darmstadt, Grafenstrasse 2, 64283 Darmstadt, Germany
}

\corrauth{Mohamed Saadi,
Faculty II - Mathematics and Natural Sciences, Technical University of Berlin,
Straße des 17. Juni 135,
Berlin,
10623, Germany}

\email{saadi@uni-wuppertal.de}

\begin{abstract}
We quantify the uncertainty of the L\"{a}mmer model of damage evolution when fitted to (noisy) observations of damage evolution in cyclic fatigue experiments with and without dwell time. 
We therefore develop a bootstrap method by sampling over blocks of load cycles in the experiments in order to quantify the uncertainty in the material parameters of the L\"{a}mmer damage evolution equation. We first develop a resilient optimization algorithm for parameter identification based on numerical solutions of damage evolution. The uncertainty is quantified on three levels: distribution of parameters of the L\"{a}mmer model, confidence bands for the solutions of damage evolution, and distributions of failure times. The method is tested on several data sets, committing considerable high-performance computing resources to the task.
\end{abstract}


\maketitle

\section{Introduction}
\label{sec:introduction}
%%%%%%%%%%%%%%%%%%%%%%%%%%%%%%%%%%%%%%%%%%%%%%%%%%%%%%%%%%%%%%%%%%%%%%%%
The reliable characterization of material durability is essential for the construction of heavily loaded mechanical components as e.g. turbine components deployed in power plants.
Future trends toward a more flexible power generation, the need to maximize the service life of high-temperature components, and the still ongoing demand to decrease conservativeness require improved and more flexible lifetime assessment methods.
More or less state of the art procedures for the assessment of component life are based on constitutive models for elasto-viscoplastic deformation in conjunction with evolutionary models for a damage variable $D(t)$, see \citet{murakami2012continuum} for instance. The damage parameter $D(t)$ increases in time until it reaches a critical value that defines failure.  Usually, the dynamics of the damage state is modeled as a differential equation driven by a function depending on the free energy and the rate of change in the stress state. As this function depends on several parameters, one has to solve an inverse problem to calibrate the damage dynamics with data, see e.g. \cite{tarantola2005inverse}. The outcome of the solution to the inverse problem relies on (a) the stability of the optimization procedure that determines the parameters and (b) the data set of (noisy) observations that is underlying the calibration procedure. 

Due to both factors, variation in the prediction of damage evolution models makes up a large part of the model uncertainty, also known as epistemic uncertainty \cite{hullermeier2021aleatoric}.   Despite the fact that aleatoric uncertainty also is a major factor in the probabilistic characterization of material properties \cite{Koelzow2019_ProbabilisticLifetimeAssessmentApproach,schmitz2013probabilistic,made2018combined}, the quantification of epistemic uncertainty is of crucial, as this uncertainty affects all instances in a fleet of devices, whereas aleatoric uncertainty only affects single instances. A cluster risk, however, has a much more severe business and safety impact as compared to a scattered risk.   

In this work, we present an approach to quantify epistemic uncertainty for damage evolution models taking the L\"{a}mmer model \cite{Laemmer1998_ThermoplastizitatUndThermoviskoplastizitatMitSchadigungBeiKleinenUndGroenDeformationen} as a suitable representative. The L\"{a}mmer model is a differential equation for the damage parameter $D(t)$ which is driven by a function of the damage state itself and the change rate of the accumulated plastic strain over time $\dot s(t)$. Further details can be found in Subsection \ref{sec:fkoelzow_damage_evolution_equations}. The L\"{a}mmer model comes with five free parameters in total, which have to be calibrated from (noisy) experimental observations. 

On the one hand, this entails the necessity to specify the observational content of the damage state $D(t)$. This can be resolved by the definition of $D$ as a decrease of stiffness, or, non-equivalently but commonly treated as, a decrease of Young's modulus $E(t)=(1-D(t))E$ for a specific reference volume element. Here, $E(t)$ is measurable from the strain-stress hysteresis of the load cycle the material runs through at time $t$ in a strain-controlled experiment, and $E$ is Young's modulus of the virgin material. 
Theoretically, it can be derived from the second law of thermodynamics \cite[eqn. 58, 120]{Reckwerth2003_Theprincipleofgeneralizedenergyequivalenceincontinuumdamagemechanics}
that $D(t)$ is monotonic increasing. 
While these clarify the observational content of the model \cite[sec. 7.2.3]{lemaitreChaboche2009_mechanicsOfSolidMaterials}, it poses new challenges as measured data do not always lead to a monotonic increase in the observational parameter $D(t)$ due to measurement artifacts. 
We are thus forced to consider this observational data as noisy.     

On the other hand, fitting the solution of an ordinary differential equation to data via parameter optimization is an inverse problem. Extensive mathematical literature exists on inverse problems, mostly associated with gradients based optimization where the gradients are computed by the adjoint method, see, e.g., again \cite{tarantola2005inverse}. In the present case, however, the ordinary differential equation for damage evolution is coupled to a constitutive model for the evolution of plasticity, namely the Chaboche model, see \cite{ambroziak2007identification} and variants thereof. As the dynamic equation of such models contain non differentiable functions, the adjoint method can not be simply applied. Instead, we resort to pseudo simplex methods \cite{singer2009nelder} for the fitting of parameters. 

Furthermore, the L\"{a}mmer model depends non-linearly on some of the parameters, which renders the optimization non-convex. This implies that multiple local minima can occur, all leading to more or less decent fitting results. In this regard, the L\"{a}mmer model is only one typical example for all kinds of damage evolution models. Non-convexity results in multi-modal parameter sets, which make it hard to consider the retrieved parameter values as 'material properties', although the damage curves themselves characterize the material to a certain precision. Furthermore, as the data contains observational noise, the uncertainty generated by this noisiness spills over to parameter and damage development curves, which needs to be properly assessed. This amounts to the computation of (multi-modal) parameter distributions or distributions of damage evolution curves.

Several approaches exist for uncertainty quantification in the context of parameter-dependent evolution equations. Stochastic Galerkin approaches, see e.g. \cite{Xiu2002_TheWienerAskeypolynomialchaosforstochasticdifferentialequations}, are very popular, however they require normal distributions as input. As we will show here, this can not be taken for granted. Also, it is not clear how this distribution can be retrieved from the data. Similarly, asymptotic maximum likelihood theory provides a normal distribution of parameter uncertainty \cite{ferguson2017course} which is computed from the data. Again, this distribution is uni-modal and, in addition, depends on a proper specification of the likelihood. The latter however is highly non-trivial in the context of observations correlated over time \cite{brockwell2009time}. Bayesian methods enjoy much popularity \cite{stuart2010inverse}, but the specification of priors and the computation of posteriors are problematic for multi-modal local optima. Also in this case, the difficulties of a proper specification of the likelihood stands in the way of computing posterior distributions.  

Here we take a different approach and resort to non parametric bootstrapping \cite{efron1982jackknife} as the method for an uncertainty quantification in the given context. This method however comes with a catch when applied to a few log-lasting experiments that produce correlated data in the time series of a single experiment. From the standpoint of material science, the evolution of the damage parameter $D(t)$ will depend on microstructural random structure.  It can be expected that this will only correlate the data over a span of load-cycles to overcome local obstacles and approximate independence is still expected to be valid over longer times. We therefore bootstrap over blocks of several hundreds of load cycles and resample on the level of blocks. By randomly recombining such blocks to a dataset on which we fit the parameters, we obtain bootstrapped parameter distributions which we subsequently analyze and use to compute the uncertainty of damage evolution curves and the associated time to failure. 

The main contribution of our paper is a methodology to asses epistemic uncertainty, and hence model quality, in the field of damage mechanics.
We also develop a high performance computing framework that is capable to actually solve the bootstrap task and test it extensively on experimental data.
From a practical engineering perspective, especially the 
uncertainty quantification and the subsequent probabilistic
results are the key to practical lifetime applications of continuum damage mechanic (CDM) models.

The paper is organized as follows: In Section \ref{sec:kzo_lifetime_assessment} we provide background information on damage evolution equations. Section \ref{sec:kzo_lcf_data_and_its_assessment} discusses the experimental setup and the data that has been acquired. Section \ref{sec:bootstrap} gives a thorough description of our bootstrap approach and in Section \ref{sec:saadi_results} we present our numerical results in terms of parameter- and curve distributions. We discuss our findings  and present our conclusions in the final Section \ref{sec:conclusion}.

%%%%%%%%%%%%%%%%%%%%%%%%%%%%%%%%%%%%%%%%%%%%%%%%%%%%%%%%%%%%%%%%%%%%%%%%%%%%%%%%
%% Author: Felix Kölzow
\section{Lifetime assessment of high-temperature components}
\label{sec:kzo_lifetime_assessment}
%%%%%%%%%%%%%%%%%%%%%%%%%%%%%%%%%%%%%%%%%%%%%%%%%%%%%%%%%%%%%%%%%%%%%%%%%%%%%%%%

%%%%%%%%%%%%%%%%%%%%%%%%%%%%%%%%%%%%
% Author: Felix Kölzow
\subsection{Overview, issues and motivation }
\label{sec:kzo_overview_issues_motivation}
%%%%%%%%%%%%%%%%%%%%%%%%%%%%%%%%%%%%

The following uniaxial viscoplastic material model in equation 
\eqref{eq:kzo_simple_viscoplastic_ode_with_damage},
see \cite{Chaboche1989_ConstitutiveEquationsforCyclicPlasticityandCyclicViscoplasticity} or \cite{ambroziak2007identification}, is used
here to enlighten the application of the damage evolution equation and to 
introduce the necessary notation:
\begin{subequations}\label{eq:kzo_simple_viscoplastic_ode_with_damage}
\begin{align}
F &= \frac{\lvert \sigma - X\rvert}{1-D} - R -k \,, \label{eq:kzo_overstress} \\
\dot s &= \Bigl\lvert \dot \varepsilon - \frac{\dot \sigma}{E}\Bigr\rvert = \lvert \dot \varepsilon_p \rvert \,, \label{eq:kzo_accumulated_plastic_strain_rate} \\
    \dot \sigma &= E\Bigl( \dot \varepsilon - \foeppl{ \frac{F}{K}}^n \sign(\sigma-X) \Bigr)\,, \label{eq:kzo_stress_rate}\\
     \dot R &= b(R_1 - R) \dot s \,,\label{eq:kzo_isotropic_hardening} \\
     % %
     \dot X &= \frac{2}{3} a(\dot \varepsilon - \frac{\dot \sigma}{E}) -cX \dot s  \,, \label{eq:kzo_kinematic_hardening} \\
     % %
     \dot D &= f(\sigma,\varepsilon_p,p)  \label{eq:kzo_damage_in_general} \,.
\end{align}
\end{subequations}

In equation \eqref{eq:kzo_overstress} the uniaxial mechanical
stress is denoted by $\sigma$, the kinematic hardening effects are modeled by the backstress $X$, while
the isotropic hardening influence is given by $R$ and 
the isotropic damage state is represented by a scalar variable denoted by $D$.
These variables determines the viscoplastic overstress $F$.
The difference in \eqref{eq:kzo_accumulated_plastic_strain_rate}
of the strain rate $\dot \varepsilon$ and the ratio of
$\dot \sigma/E$ represents the plastic strain rate $\dot \varepsilon_p$, while 
this relationship is valid under the small-strain assumption. In the context of an uniaxial loading scenario, 
the rate of the accumulated plastic strain $\dot s$ is simply the absolute value of
the plastic strain rate $\dot \varepsilon_p$. The 
equation \eqref{eq:kzo_isotropic_hardening} and \eqref{eq:kzo_kinematic_hardening} concern
the evolution of the isotropic and kinematic hardening behavior, respectively. 
The model parameters regarding
the viscoplastic material behavior are $n,b,R_1$, the initial yield stress $k$ 
and the Young's modulus of the virgin material $E$.
The last equation is a more general one to represent different damage evolution equations,
while the evolution depends on $\sigma$, $\varepsilon_p$ and further damage model
parameters $p$. 
The equation
\eqref{eq:kzo_damage_in_general} is of particular interest,
since it (should) allows for the lifetime prediction under cyclic loading and high-temperature conditions.

In order to characterize the reliability and the lifetime of turbine components
in power plants under typical loading scenarios, i.e.  cyclic loading and high-temperature conditions, the equation \eqref{eq:kzo_damage_in_general} is of particular interest. In principle, equation \eqref{eq:kzo_damage_in_general} should be suitable for predicting the service life under the mentioned loading scenarios.

% paragraph
But there are also different methods regarding the lifetime computation of high-temperature components were developed for
decades, see e.g. \cite{danzer1988_lebensdauerprognoseHochfesterMetallischerWerkstoffeImBereichHoherTemperaturenOCR} for a general overview. 
Common to many lifetime methods is a predefined summation rule whereby mechanical stress and strain
conditions and load case parameters such as temperature are used to determine the current damage state $D(t)$.
These lifetime methods are also called damage accumulation methods since they accumulate damage increments based on summation rules. 
A representative example is termed 
the \textit{generalized damage accumulation rule} described 
by \cite{scholzBerger2005_deformationLifeAssessmentOfHightemperatureMaterialsUnderCreepFatigueLoading}.

% paragraph
Some lifetime methods were written into guidelines to provide solid foundations for a more unified lifetime assessment. An overview and comparison of selected codes and standards are summarized by  
\cite{Schmitt2017_LifeAssessmentMethodsComparisonMPASeminar2017.pdf}, while 
\cite{Takahashi2013_SystematicEvaluationofCreep-FatigueLifePredictionMethodsforVariousAlloys} gives a 
methodological comparison. Especially in the latter publication, it is stressed that the critical damage 
threshold $D_\mathrm{crit}$, which 
represents a relationship between damage and the time to technical crack initiation
$t_\mathrm{I}$ with $D(t_I)=D_\mathrm{crit}$,
differs dramatically between different methods and materials.

% paragraph
Considering reliability and safety requirements, a methodological balance between more sophisticated lifetime methods and the possibility of quantifying safety 
and reliability had and have to be found. The combination of a damage accumulation method
and state-of-the-art reliability methods were shown in
\cite{harlowDelph1997_probabilisticModelForCreepFatigueFailure,maoMahadevan2000_reliabiltyAnalysisOfCreepFatigueFailure,
Koelzow2017_ProbabilisticLifetimeAssessmentinCaseofCreepFatigueInteractionofHighTemperatureMaterials, Koelzow2019_ProbabilisticLifetimeAssessmentApproach}. It turned out that the simple structure of damage accumulation models 
lead to relatively simple lifetime predictions but makes it even harder to apply sophisticated probabilistic methods. 

% paragraph
Furthermore, the application of the CDM, for example given by \cite{lemaitreChaboche2009_mechanicsOfSolidMaterials} or \cite{murakami2012continuum}, combined with sophisticated probabilistic methods promises a more concise model structure when complex loading scenarios are considered, see for instance \cite{Koelzow2020_APPLICATIONOFDAMAGEMECHANICSANDPOLYNOMIALCHAOSEXPANSION} as well as \cite{tuprints18616}.

%%%%%%%%%%%%%%%%%%%%%%%%%%%%%%%%%%%%
% Author: Felix Kölzow
\subsection{Damage evolution equations}
\label{sec:fkoelzow_damage_evolution_equations}
%%%%%%%%%%%%%%%%%%%%%%%%%%%%%%%%%%%%

The most challenging part of lifetime prediction using the CDM is choosing the damage evolution equation. Several approaches exist, and an incomplete list is given in \cite{tuprints187}.

% paragraph
A combination of damage evolution equations proposed by 
\cite{WEIHUATAI1986377} and \cite{lemaitreChaboche2009_mechanicsOfSolidMaterials}, while
an additional term that models the void nucleation, leads to the damage evolution equations given by \cite{Dhar1996_Acontinuumdamagemechanicsmodelforvoidgrowthandmicrocrackinitiation}.
A further extension, introduced by \cite{Laemmer1998_ThermoplastizitatUndThermoviskoplastizitatMitSchadigungBeiKleinenUndGroenDeformationen}, reads:
%
% NOTE: \vphantom{} is used in order to achieve the same height for
% the curly brackets generated by the underbrace command
\begin{align}\label{eq:kzo_laemmer_damage}
  \dot D = \underbrace{\vphantom{\frac{(-Y)^{\mycolor{q_0}}}{(1-D)^{\mycolor{q_1}}}}\qquad \mycolor{\alpha_0} \dot s\qquad }_{\text{void nucleation}} + \underbrace{(\mycolor{\alpha_1} + \mycolor{\alpha_2} D) \frac{(-Y)^{\mycolor{q_0}}}{(1-D)^{\mycolor{q_1}}} \dot s}_{\text{void growth}} \,.
\end{align}
The relationship \eqref{eq:kzo_laemmer_damage} contains f\/ive temperature-dependent material parameters $\alpha_0,\alpha_1,\alpha_2,q_0,q_1 \in \mathbb{R}_0^+$.

% paragraph
The damage energy release rate density $Y$, cmp. \cite{Saanouni1994_OntheAnelasticFlowwithDamage},
can be obtained as a derivative of a free energy function $\Psi$ with respect to the damage variable $D$ and reads
\begin{align}
\label{eq:Y_from_Psi}
  -Y = \varrho \frac{\partial \Psi}{\partial D}\,,
\end{align}
while $\varrho$ is the material density. Additionally, 
$\dot s$ denotes the
accumulative plastic strain rate, which can be obtained by
the plastic part of the strain rate tensor $\dot E^P_{ij}$, for more details see \cite{lemaitreChaboche2009_mechanicsOfSolidMaterials}:
\begin{align}
  \dot s = \sqrt{ \frac{2}{3} \dot E^p_{ij}\dot E^p_{ij} }\,,
   \text{ uniaxial case:}\quad \dot s = \lvert\dot \varepsilon^p\rvert \,.
\end{align}

On the one hand, there is a part of the literature, 
for example \cite{Sermage2001_Multiaxialcreepfatigueunderanisothermalconditions},
and ~\cite{lemaitreChaboche2009_mechanicsOfSolidMaterials},
which uses only the elastic part of $\Psi$, i.e. neglecting
the kinematic and isotropic energy parts since it is assumed, as stated
in \cite[p. 400]{lemaitreChaboche2009_mechanicsOfSolidMaterials}, 
that hardening and other effects are decoupled from elasticity and damage. 
On the other hand,
e.g. \cite{Saanouni1994_OntheAnelasticFlowwithDamage},
\cite{Reckwerth2003_Theprincipleofgeneralizedenergyequivalenceincontinuumdamagemechanics}
and \cite{samir2006_konsitutiveWerkstoffbeschreibungKriechermuedung},
consider the complete term of $\Psi$, i.e. including the kinematic and isotropic energy parts.

We consider only the elastic energy component and obtain the following expression for $Y$ under uniaxial loading:
\begin{align}
\label{eq:elastic_Y}
  -Y(t) = \frac{1}{2} \frac{\bigl( \sigma(t)\bigr)^2}{(1-D(t))^2E}\,.
\end{align}

The methodological advantage behind the assumption of decoupled 
damage and hardening effects is enlightened in equation \eqref{eq:elastic_Y}.
The damage energy release rate density $Y$ is governed  
by the uniaxial stress $\sigma(t)$ and the current damage state $D(t)$ 
which both can be measured indirectly during a strain- or stress-controlled
low-cycle fatigue experiment. There is no use of any internal variable
that further describes the internal material state and thus
is hidden even to indirect experimental measurement. As a consequence,
equation \eqref{eq:elastic_Y} is valid for different viscoplastic 
material models and even further, directly allows the usage
of the indirectly measured mechanical stresses and the current
damage state to determine $Y$. Moreover, the statistically inferred
damage model parameters in equation \eqref{eq:kzo_laemmer_damage}
can be universally used for different viscoplastic material models.
The described methodological advantage and 
its simpler
applicability compared to the assumption of coupled
hardening and damage effects does not mean 
that the uncoupled behavior is or was experimentally corroborated. 

\begin{figure}[t]
    \centering
  \includegraphics[width=6.5cm]{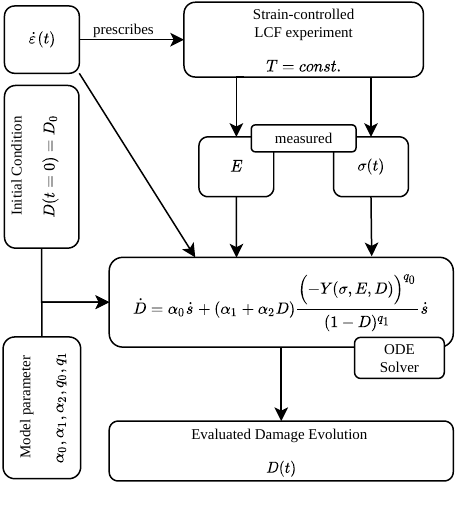}
  \caption{
    Input parameters and the material response are used to
    compute the damage evolution for a low-cycle fatigue experiment. 
    No additional model regarding the deformation behavior is necessary
    for this procedure. This is possible, if
    and only if the elastic part $\Psi_e$ of the free energy function is used to
    compute $Y$ according to \eqref{eq:Y_from_Psi}.
    }
  \label{fig:kzo_how_do_compute_damage_evolutions}
\end{figure}

In order to evaluate the damage evolution for a specific damage model equation,
only the mechanical stress $\sigma(t)$, the strain rate $\dot \varepsilon(t)$, the Young's modulus $E$ of the undamaged material,
the damage model parameters and initial condition $D(t=0) = D_0$ have to be known to compute the damage evolution over time, $D(t)$.
The interaction of different model components is shown in figure~\ref{fig:kzo_how_do_compute_damage_evolutions}.
%

% paragraph
Finally, the end-of-life for a specific reference volume element or more 
technically, the number of cycles until a technical crack initiation is 
reached, can be defined such that $D(t=t_{I}) = D_{\mathrm{crit}}$ is met.
The critical damage threshold is denoted by $D_\mathrm{crit}$ while $t_I$ represents
the time until technical crack initiation is observable.
The critical damage threshold has to be determined experimentally, and as
suggest by \cite{tuprints18616},
a value of \linebreak $D_\mathrm{crit} = 0.1$ is a reasonable one.
Reasonable in this case means that a value of $D_\mathrm{crit} = 0.1$
correlates with the conventional load drop criterion of five percent, see 
e.g. \cite[sec. 7.9]{ISO12106}.

%%%%%%%%%%%%%%%%%%%%%%%%%%%%%%%%%%%%%%%%%%%%%%%%%%%%%%%%%%%%%%%%%%%%%%%%%%%%%%%%
% Author: Felix Kölzow
\section{Experimental low-cycle fatigue data}
\label{sec:kzo_lcf_data_and_its_assessment}
%%%%%%%%%%%%%%%%%%%%%%%%%%%%%%%%%%%%%%%%%%%%%%%%%%%%%%%%%%%%%%%%%%%%%%%%%%%%%%%%
In this section we discuss parameter identification for CDM and uncertainty quantification of parameters. 
%%%%%%%%%%%%%%%%%%%%%%%%%%%%%%%%%%%%
\subsection{Data sets overview and its origin }
\label{sec:kzo_data_set_origin}
%%%%%%%%%%%%%%%%%%%%%%%%%%%%%%%%%%%%

If a specific model or a model family of damage evolution equations is selected,
the corresponding model parameters have to be inferred from experimental data. 
The data can be obtained by strain-controlled creep-fatigue tests 
at high-temperature conditions. For decades, many of those
creep-fatigue tests were performed at the Department and Institute of Materials Science (IfW) at the Technical University of Darmstadt.

\begin{figure}[h]
  \centering
  \begin{tikzpicture}
  [scale=1,z=-0.8cm,thick,axis/.style={thin, ->, >=stealth'}]
  % 
  % define general coordinates
  % 
  \coordinate  (ursprung) at (-3,0);
  \coordinate (endx1) at (2,0) {};
  \coordinate (endx2) at ( $(ursprung) + (0,1.5) $);
  % 
  % fixed coordinate system 
  % 
  \draw [axis] ($(ursprung) - (0.5,0)$) -- (endx1) node (x1line) [right]{$t$}; % x1-achse
  \draw [axis] ($(ursprung) -(0,1.125)$)  -- (endx2) node (x2line)
  [below left]{$\varepsilon(t)$}; % x3-achse
  % 
  % 
  % 
  % MORE  coordinates
  % 
  \coordinate (startCycle) at ($(ursprung) + (1.0,0)$);

  \coordinate (rampup) at (0.4,1.0);
  \coordinate (rampdown) at (0.4,-1.0);
  \coordinate (dwelltime) at (1,0);
  \coordinate (strainrange) at (0,2.0);

  \coordinate (thOneLeft) at ($(startCycle) + (rampup)$);
  \coordinate (thOneRight) at ($(thOneLeft) + (dwelltime)$);

  \coordinate (thTwoLeft) at ($(thOneRight) + (rampdown) + (rampdown)$);
  \coordinate (thTwoRight) at ($(thTwoLeft) + (dwelltime)$);

  \coordinate (endCycle) at ($ (thTwoRight) + (rampup)$);

  % draw cycle
  \draw [solid] (startCycle) -- (thOneLeft) -- (thOneRight) -- (thTwoLeft) -- (thTwoRight) -- (endCycle);

  %% measurement stuff
  % strain range
  \coordinate (thOneCenter) at ($0.5*(thOneRight) - 0.5*(thOneLeft) + (thOneLeft) $);
  \coordinate (thOneCenterAbove) at ($(thOneCenter) - (strainrange)$);
  \draw[<->,>=stealth,very thin] (thOneCenter) -- (thOneCenterAbove);
  \draw[very thin] (thTwoLeft) -- ($(thOneCenterAbove) + (-0.1,0)$);
  \node[rotate=90,font=\footnotesize,above] at ($(thOneCenter) - 0.75*(strainrange)$) {$\Delta \varepsilon$};

  % max strain
  \draw[very thin, <->] ($(-0.25,0) + (startCycle)$) --  ++($0.5*(strainrange)$);
   \draw[very thin] (thOneLeft)  -- ($(-0.25,0) + (startCycle) + 0.5*(strainrange) + (-0.1,0)$);
 \node[font=\footnotesize,rotate=90,above] at ($(startCycle)  + (-0.25,0) + 0.25*(strainrange)$) {$\varepsilon_{\max}$};
    
  % min strain
  \draw[very thin,<->] ($(thOneRight) - (strainrange)$) -- ($(thOneRight) - 0.5*(strainrange)$);
  \node[rotate=90,font=\footnotesize,above] at ($(thOneRight) - 0.75*(strainrange)$) {$\varepsilon_{\min}$};
  
  %% thOnce
  \coordinate (helperDistance) at (0,0.3);
  \coordinate (offset) at (0,0.1);
  \coordinate (thOneLeftHelper) at ($(thOneLeft) + (helperDistance)$);
  \coordinate(thOneRightHelper) at ($(thOneRight) + (helperDistance)$);
  \draw[very thin] (thOneRight) -- ($(thOneRightHelper) + (offset)$);
  \draw[very thin] (thOneLeft) -- ($(thOneLeftHelper)+ (offset)$);
  \draw[<->,>=stealth,very thin] (thOneLeftHelper) -- (thOneRightHelper);
  \node[font=\footnotesize] at ($ 0.5*(thOneRightHelper) - 0.5*(thOneLeftHelper) + (thOneLeftHelper) + (0,0.2)$) {$t_{d,t}$};

  %% thTwo
  \coordinate (thTwoLeftHelper) at ($(thTwoLeft) + (helperDistance)$);
  \coordinate(thTwoRightHelper) at ($(thTwoRight) + (helperDistance)$);
  \draw[very thin] (thTwoRight) -- ($(thTwoRightHelper) + (offset)$);
  \draw[very thin] (thTwoLeft) -- ($(thTwoLeftHelper)+ (offset)$);
  \draw[<->,>=stealth,very thin] (thTwoLeftHelper) -- (thTwoRightHelper);
  \node[font=\footnotesize] at ($ 0.5*(thTwoRightHelper) - 0.5*(thTwoLeftHelper) + (thTwoLeftHelper) + (0,0.2)$) {$t_{d,c}$};

  %% Point annotations
  \coordinate (mypoint) at ($ (thOneRight) +  0.5*(rampdown)$);
  \coordinate (annotationPoint) at ($ (mypoint) +  (0.25,0.25)$);
  % left point
  \draw[very thin,dashed] ($(startCycle) + 0.5*(rampup)$) -- (annotationPoint); % -- ($(annotationPoint) - (1.3,0) $) -- (annotationPoint);
  \fill ($(startCycle) + 0.5*(rampup)$) circle[radius=1.0pt];
  % center point
  \draw[very thin,dashed] (mypoint) -- (annotationPoint) node[above right, font=\footnotesize] {$\dot \varepsilon = \textrm{const.}$};
  \fill (mypoint) circle[radius=1.0pt];
  % right point
  \draw[very thin,dashed] ($(endCycle) - 0.5*(rampup)$) -- (annotationPoint);
  \fill ($(endCycle) - 0.5*(rampup)$)  circle[radius=1.0pt];

\end{tikzpicture}

%%% Local Variables:
%%% mode: latex
%%% TeX-master: "../main_publication"
%%% End:
  \caption[Service-type (steam turbine) cycle.]
  {
  Trapezoidal cycle. - $\varepsilon(t)$ strain, $\dot \varepsilon$ strain rate,
  $\Delta \varepsilon$ strain range,
  $\varepsilon_{\min}$ min. strain amplitude,
  $\varepsilon_{\max}$ max. strain amplitude,
  $t_{d,t}$ dwell time in tension,
  $t_{d,c}$ dwell time in compression.
  }
\label{fig:ifw_trapezoidal_cycle}
\end{figure}
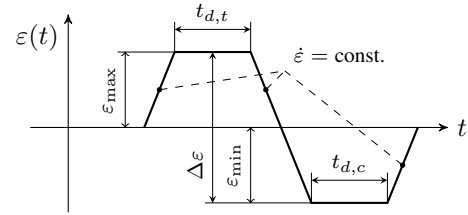

% paragraph
The strain-controlled low-cycle fatigue experiments were carried out in different research projects as described in \cite{tubiblio85203} and \cite{FVV1259_AVIF307_DWG2_jobski}.
While several creep-fatigue experiments at different temperature levels are available regarding 
the investigated high-chromium cast steel {GX13CrMoCoVNbNB9-2-1},
we focus on data recorded at \SI{600}{\celsius} which can be considered as 
main temperature for this kind of material. The strain-time function, used to describe the load 
case for each low-cycle fatigue experiment, constitutes a trapezoidal or triangular cycle form.
The trapezoidal cycle form is shown figure \ref{fig:ifw_trapezoidal_cycle}. 

This cycle form consists of strain ramps up and down with constant strain 
rate $\dot \varepsilon$ and a dwell time $t_{d,t}$ under tensile loading 
and a further dwell time $t_{d,c}$ under compression. The strain range $\Delta \varepsilon$ 
determines the difference between the maximum and
minimum applied strain amplitude, i.e. $\Delta \varepsilon = \varepsilon_{\max} - \varepsilon_{\min}$. 
As a special case, the triangular cycle form appears if the dwell times are set to zero. 
This both forms of cycling loading are continued until the end-of-life criterion 
is reached and this declares the end of the LCF experiment. 

The low-cycle fatigue and creep-fatigue experiments selected for this article are summarized
in table \ref{tab:kzo_lcf_experiments_overview}. The first three characters within 
the \textit{specimen-id} declares a specific heat of the material 
{GX13CrMoCoVNbNB9-2-1}, the integer 6 indicates the applied isothermal 
temperature at \SI{600}{\celsius} and the abbreviation
\textit{d} or \textit{dh} determines whether it is a 
triangular cycle type or trapezoidal cycle type, respectively. 

\begin{table*}
\centering
 \renewcommand{\arraystretch}{1.5}  
\begin{tabular}{cccc>{\centering\arraybackslash}m{1in}}
\toprule
 Specimen-Id       & $\Delta \varepsilon$\, / $(-) / 10^2$ & $t_{d,t}$ / $\SI{}{\min}$ & $t_{d,c}$ / $\SI{}{\min}$ & Cycle type \\
\midrule
BDE6d1  & 2                         & 0         & 0                        &   \includegraphics[keepaspectratio,width=8mm]{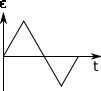}   \\
BDE6d2  & 1                         & 0         & 0                        &   \includegraphics[keepaspectratio,width=8mm]{pictures/triangularCycle.png}    \\
BDE6d3  & 0.55                      & 0         & 0                       &    \includegraphics[keepaspectratio,width=8mm]{pictures/triangularCycle.png}   \\
BDE6d4  & 0.35                      & 0         & 0                        &    \includegraphics[keepaspectratio,width=8mm]{pictures/triangularCycle.png}   \\
BDE6dh1 & 1                         & 3         & 3                        &   \includegraphics[keepaspectratio,width=8mm]{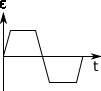}    \\
BDE6dh2 & 0.55                      & 3         & 3                        &   \includegraphics[keepaspectratio,width=8mm]{pictures/trapezoidalCycle.png}    \\
BDE6dh3 & 0.55                      & 10        & 10                      &    \includegraphics[keepaspectratio,width=8mm]{pictures/trapezoidalCycle.png} \\
\bottomrule
\end{tabular}
\caption{Overview of the experimental data selected. The table lists 
 the strain range $\Delta \varepsilon$, the dwell time in tension $t_{d,t}$ as well as compression $t_{d,c}$, the strain ratio $R_{\varepsilon}$ and the type of cyclic loading.
 The strain ratio for all investigated experiments is 
 $R_{\varepsilon} = \varepsilon_{\min} / \varepsilon_{\max} = -1$. 
  }
 \label{tab:kzo_lcf_experiments_overview}
\end{table*}

%%%%%%%%%%%%%%%%%%%%%%%%%%%%%%%%%%%%
% Author: Felix Kölzow
\subsection{Experimental damage evaluation}
\label{sec:kzo_data_and_damage_assessment}
%%%%%%%%%%%%%%%%%%%%%%%%%%%%%%%%%%%%

The concept of CDM allows a definition of material 
damage and different possibilities
to quantify its evolution. Classical methods regarding the 
damage measurement are described in \cite{Yuan-Sheng1988_Measurementofcontinuousdamageparameter} and \cite{lemaitreChaboche2009_mechanicsOfSolidMaterials}. 
One possible way to determine the damage evolution is based on
the stiffness reduction of a reference volume element or, in
the present context, the stiffness reduction of the specimen.
Since the number and dimension of micro-cracks and micro-cavities are expected to increase over time, 
the corresponding stiffness decreases over time. For that reason,
determining the Young's modulus at different time points
during the low-cycle fatigue experiment is a possible way
to quantify the damage evolution.
Each measured hysteresis loop of the stress-strain response at high-temperature
conditions offer the possibility of measuring the Young's modulus.

% paragraph
If the Young's modulus $E_{T}(t_i)$ and the corresponding point in time $t_{i}$ 
is determined for every measured hysteresis loop $i$, the measured damage evolution $D(t_i)$
is given by (see e.g. \cite{lemaitreChaboche2009_mechanicsOfSolidMaterials})
\begin{align}
 D(t_i) = 1 - \frac{E_{T}(t_i)}{E} \,,
\end{align}
where $E$ specifies the reference Young's modulus of the virgin material. 
The index $T$ in $E_T$ refers, in accordance to
\cite[p. 4]{ISO12106}, to the Young’s modulus determined from
the unloading branch of the hysteresis loop. 

% paragraph
As an approximation, it is assumed
that the specimen at the beginning of cyclic loading is free of 
damage and no significant damage
is induced during the first $N_0 = 10$ cycles. As a consequence, 
we define the reference $E$
as an average of $E_T(t_i)$ for the first $i=1,\ldots, N_0$ cycles, i.e.  $E = \frac{1}{N_0} \sum^{N_0}_{i=1} E_T(t_i)$.

\begin{figure}[b] 
  \centering
  
\pgfplotsset{
every axis legend/.append style={
at={(0.025,0.125)},
anchor= south west },
tick label style={font=\small},
label style={font=\small},
legend style={font=\small},
title style={font=\small},
compat=1.11
}

\begin{tikzpicture}
  [axis/.style={thick, ->, >=stealth'}]

  \begin{axis}
    [legend cell align=left,
    /pgf/number format/.cd,
    fixed,precision=2, zerofill={true}, % show zeros 
    width=70mm,
    height=70mm,
    grid=both,
    xlabel = {normalised strain},
    ylabel = {normalised stress},
    xtick={-1.0,-0.5,...,1.0},
    xmin=-1.2,
    xmin=0,
    xmax=1.2,
    ymin=0,
    ymax=1.5,
    ytick={-1.0,-0.5,...,1.0},
    ylabel near ticks,
    ]
    \pgfplotstableread[col sep = comma]
    {data/normalized_hysteresis_loop.csv}\datatable;
    \addplot[mark size = 1.0,black,solid,gray,very thick] table[x=strain, y=stress] from \datatable;
    \addlegendentry{Hysteresis};
       % epsilon_l
    \draw[-stealth] (0.0,1.1)--(0.875,1.1); 
    \draw (0.875,1.125) -- (0.875,0.725);
    \node at (0.40,1.1) [above]{$\varepsilon_l$};
    \fill (0.875,0.725) circle[radius=0.75pt];
    % epsilon_u
    \draw[-stealth] (0.0,1.225)--(0.975,1.225); 
    \draw (0.975,1.25) -- (0.975,0.935);
    \node at (0.40,1.225) [above]{$\varepsilon_u$};
    \fill (0.975,0.935) circle[radius=0.75pt];
    % epsilon_max
    \draw[-stealth] (0.0,1.35)--(1.0,1.35); 
    \draw (1.0,1.3725) -- (1.0,1.0);
    \node at (0.40,1.35) [above]{$\varepsilon_{\max}$};
    \fill (1,1) circle[radius=0.75pt];
    \draw[thin,dashed] (0,-1.1125) -- (2, 3.0875);
    \addlegendimage{thin,dashed}
    \addlegendentry{$E_T(t_i)$}; 
\end{axis}
\end{tikzpicture}
    \caption{
      Experimental determination of $E_{T}(t_i)$ where
      the unloading branch from tension was used.
      In order to determine $E_{T}(t_i)$, data points within the
      prescribed strain interval $(\varepsilon_l, \varepsilon_u)$
      are taken into account with
      $\varepsilon_l := 0.875\cdot\varepsilon_{\max}$ and $\varepsilon_u := 0.975\cdot\varepsilon_{\max}$.
    }
    \label{fig:kzo_measurementYoungsModulus}
\end{figure}
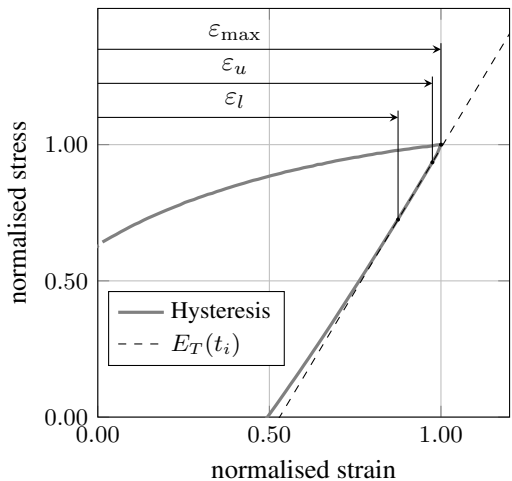

The determination of Young’s modulus during the unloading branch of a strain-controlled low-cycle fatigue 
experiment without dwell time is shown in figure \ref{fig:kzo_measurementYoungsModulus}.
To omit effects that appear due to the load reversal, a specific
strain interval is defined to determine Young’s modulus.
The predefined
strain interval, as given in figure \ref{fig:kzo_measurementYoungsModulus}, is kept unchanged for the complete evaluation 
of the data and every data set, which is recommend
\cite[p. 22]{Lemaitre1996_ACourseonDamageMechanics}.
We expect, for maximum tensile loading, that micro-cracks and micro-
cavities are open, and the maximum effective load-carrying area reduction occurs.
The corresponding damage state is called \textit{active}.
Therefore, the unloading branch under tensile loading is used to measure 
the stiffness reduction and, derived from this, the damage evolution.
In the case of maximum compressive loading conditions during the experiment,
the micro-cracks and micro-cavities are still present but are (almost) closed. Thus, no reduction in the stiffness behavior is expected, and the
damage evolution cannot be measured. The damage state is called \textit{passive}. 
This relationship is also called damage deactivation or damage-induced anisotropy
\cite{Chaboche1992_DamageInducedAnisotropy, Hansen1995_DamageDeactivation, Altenbach2001_ModellingofCreepDamageundertheReversedStressStates, Ganczarski2010_ContinuousDamageDeactivationinModelingofCycleFatigueofEngineeringMaterials}.

The individual data sets with its proprietary data
format were processed to obtain a unified
data format that simplifies data access during the 
bootstrap method explained in section \ref{sec:bootstrap}.

%%%%%%%%%%%%%%%%%%%%%%%%%%%%%%%%%%%%%%%%%%%%%%%%%%%%%%%%%%%%%%%%%%%%%%%%%%%%%%%%
% Author: Mohamed
\section{Damage parameter identification and bootstrapping}
\label{sec:bootstrap}
%%%%%%%%%%%%%%%%%%%%%%%%%%%%%%%%%%%%%%%%%%%%%%%%%%%%%%%%%%%%%%%%%%%%%%%%%%%%%%%%

\subsection{Parameter estimation}
The identification of parameters in a CDM is to calibrate the constitutive Lämmer model presented in equation \eqref{eq:kzo_laemmer_damage} with the help of experimental data. The goal is, to estimate material parameters that allow predictions within a certain range of application.\\
We need to define the cost function $F$ as the $L_2-$norm $\|\cdot\|_2$ of the residual $r$ :

\begin{subequations}
    \begin{align}
F(p) &= \|r(p)\|_2^2 = \|D^{com} - D^{exp}\|_2^2 \\ 
     &= \sum_{i=1}^{N}( D_{i}^{com}(p)-D_{i}^{exp})^2 \,,
\end{align}
\end{subequations}
where $D^{com}$ is the computational damage, $D^{exp}$ is the experimental damage and $p=\{\alpha_{0},\alpha_{1},\alpha_{2},q_{0},q_{1}\}$ is a vector of Lämmer damage parameters. The parameter identification problem can be expressed as the following constrained optimization problem:
\begin{subequations}
  \begin{alignat}{2}
    &\!\min_{p\in \mathbb{R}^5 }     &\qquad&   F(p)\,, \\
    &\text{subject to} &      & \mathrm{LB}\leq p \leq \mathrm{UB}\,,
      \end{alignat}
\end{subequations}
where
\begin{subequations}
  \begin{alignat*}{2}
    &\mathrm{LB} \quad=& &\{\alpha_{0,{min}},\alpha_{1,{min}},\alpha_{2,{min}},q_{0,{min}},q_{1,{min}}\}\, ~~~ \text{and}  \\
    &\mathrm{UB} \quad=& &\{\alpha_{0,{max}},\alpha_{1,{max}},\alpha_{2,{max}},q_{0,{max}},q_{1,{max}}\}
  \end{alignat*}
\end{subequations}
are the lower and upper bounds of the allowed Lämmer damage parameters during the optimization.

The optimization routine for fitting the damage parameters\linebreak $p=(\alpha_0,
\alpha_1,\alpha_2,q_0,q_1)$ by using Lämmer's damage evolution equation under 
consideration of a critical damage threshold of $D_\mathrm{crit} = 0.10$ has been 
developed and implemented
within the programming language for statistics, \texttt{R}. 
To solve the ordinary differential equation (ODE) presented in 
(\ref{eq:kzo_laemmer_damage}) we use the \texttt{R}-function \texttt{lsoda}, 
which is suitable for stiff or non-stiff systems of differential equations.

A gradient free optimization algorithm like Nelder-Mead are used to overcome the non-smoothness of the cost function. 
This approach is based on a pattern search method that compares values of the cost
function at the vertices of a simplex. Moreover, this approach is able to optimize cost functions without the need to compute derivatives.

There are several \texttt{R}-packages which use the Nelder-Mead algorithm to solve optimization problems. We use the R-package \textit{nloptr} that provides an \texttt{R}-interface to \texttt{NLopt}, which is a free and open-source library for nonlinear optimization \citep{Ypjo}. 

Note that a good choice of a set of initial parameters based on material expert opinions accelerates the convergence process of the optimization.

\begin{figure*}[h!]
\centering
    \includegraphics[scale=.54]{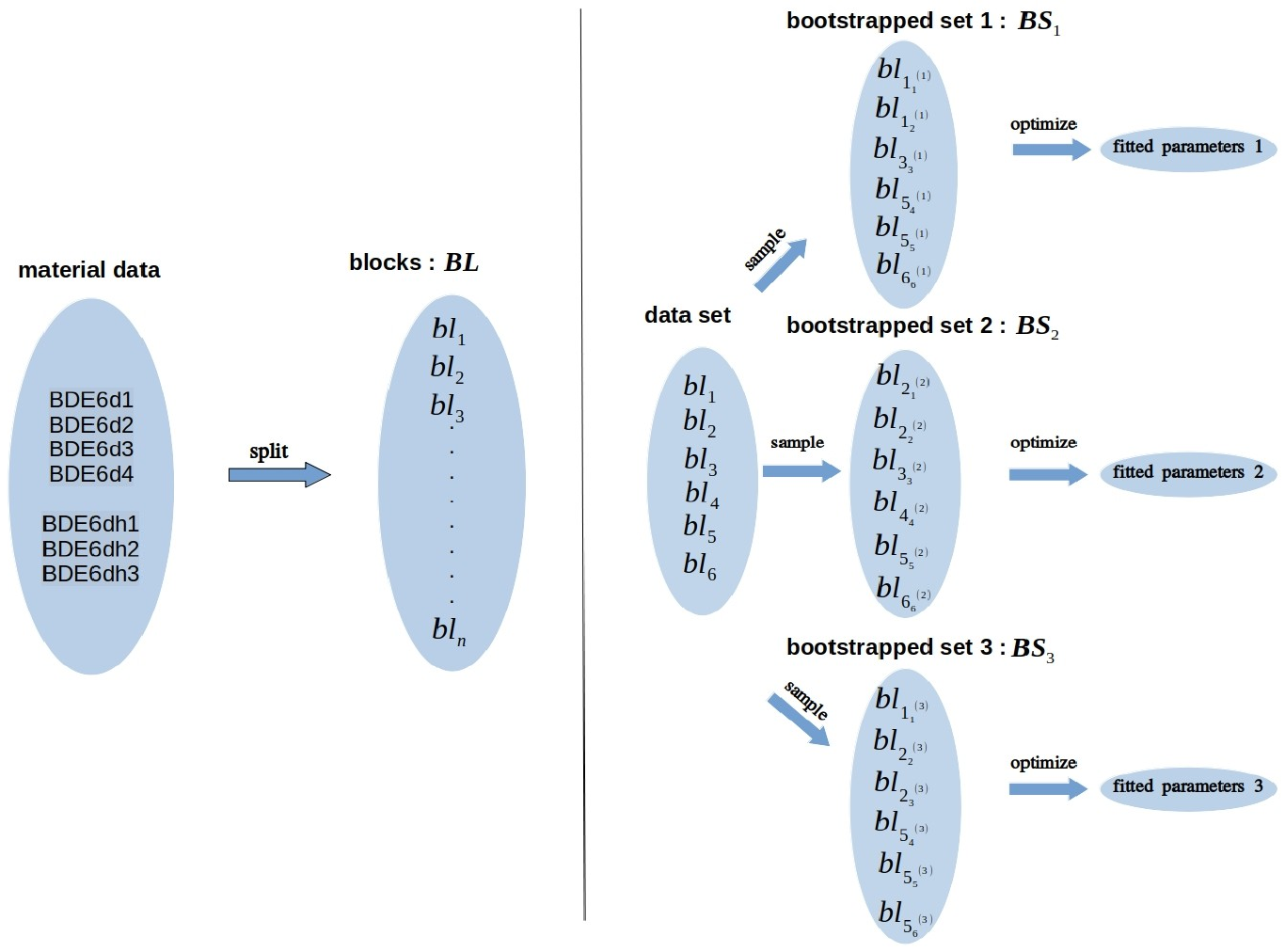}
    \caption{ The bootstrap method by sampling over blocks.
    }
    \label{fig:saadi_bootstrapping}
\end{figure*}

%%%%%%%%%%%%%%%%%%%%%%%%%%%%%%%%
\subsection{The bootstrap method}
%%%%%%%%%%%%%%%%%%%%%%%%%%%%%%%%

The second main goal besides the identification of  material parameters capable to predict the damage evolution for low-cycle fatigue experiments with different strain ranges and dwell times is  the capability to quantify the uncertainty of the CDM.
The bootstrapping technique is used to obtain improved estimates and confidence intervals for material parameters. 

% paragraph
The bootstrap, first introduced by  \cite{Efron1979_BootstrapMethodsAnotherLookattheJackknife}
and even more comprehensive in \cite{Efron1994_AnIntroductiontotheBootstrap},
is a flexible, powerful and recently developed data-based simulation method
that uses random sampling with replacement for statistical inference. 
The primary use of bootstrapping is to quantify and
assess the uncertainty associated with a given estimator without making assumptions 
about the data distribution. Further references regarding block bootstrap methods can be found in 
\cite{Lahiri1999_TheoreticalComparisonsofBlockBootstrapMethods,
Lahiri2003_ResamplingMethodsforDependentData,
Mudelsee2010_ClimateTimeSeriesAnalysis,
Lodhi2011_BootstrappingParameterEstimationinDynamicSystems,
Chernick2014_AnIntroductiontoBootstrapMethodswithApplicationstoR}
and 
\cite{Pilavakis2017_MovingBlockandTaperedBlockBootstrapforFunctionalTimeSerieswithanApplicationtotheKSampleMeanProblem}.

% paragraph
The developed bootstrap procedure is drawn in figure \ref{fig:saadi_bootstrapping}.
The size of one data set is defined as the sum of the number of 
cycles of all considered experiments. The first 
step is to split the experiment data into $N$-blocks 
$$\mathbf{BL} = \{\mathrm{bl}_1,\mathrm{bl}_2,..,\mathrm{bl}_N\}$$ with approximately the 
same size. Note that
the size of blocks can depend on the computational resources.  
Typical values of block size range from 100 to 400 cycles (see appendix \ref{sec:appendix_A1}). 

A bootstrap sample or virtual experiment $\mathrm{BS}_j,~1\leq j \leq L$ is generated from the set of original blocks $\mathbf{BL}$, while $L$ is the number of desired bootstrap samples. Let $i_l^{(j)}\in \{1,\ldots,N\}$ be independent and identically uniform distributed random variables.
Each bootstrap sample (see figure \ref{fig:saadi_bootstrapping}) 
\begin{align}
\mathrm{BS}_j = \{\mathrm{bl}_{{i}^{(j)}_{1}},\mathrm{bl}_{{i}^{(j)}_{2}},..,\mathrm{bl}_{{i}^{(j)}_{N}}\}  
\end{align}
has $N$ elements generated by sampling with replacement $N$ times from the set of original blocks $\mathbf{BL}$.
This process will be repeated $L$ times and we obtain the set of virtual experimental data that consists of bootstrapped sets:
\begin{align}
  \{\mathrm{BS}_1,\mathrm{BS}_2,...,\mathrm{BS}_L\} \,.
\end{align}
The goal of the optimization method is to identify material parameters governing the
damage evolution equation 
based on bootstrapped sets generated from the original experimental low-cycle 
fatigue data.
We define the cost function $F_{{i}^{(j)}_{r}}$ of the block $\mathrm{bl}_{{i}^{(j)}_{r}}$ as the sum of the squares of the difference between the computed damage $D_{{{i}^{(j)}_{r}}}^{com}$ and the experimental damage $D_{{{i}^{(j)}_{r}}}^{exp}$:
\begin{align}
F_{{i}^{(j)}_{r}}(p) = \sum_{k=1}^{l_{{i}^{(j)}_{r}}} ( D_{{{i}^{(j)}_{r}},k}^{com}(p)-D_{{{i}^{(j)}_{r}},k}^{exp})^2  
\end{align}
where $p=\{\alpha_{0},\alpha_{1},\alpha_{2},q_{0},q_{1}\}\in \mathbb{R}^5$ 
is the vector of Lämmer damage parameters
to be estimated and $l_{{i}^{(j)}_{r}}$ is the length of block $bl_{{{i}^{(j)}_{r}}}$. 
We note that for every block $bl_j$ we compute the cost function $F_j$. 

The generalized cost function (GCF) is defined as the sum of all cost functions computed for each block:
\begin{align}
  \mathrm{GCF}(p)= \sum_{{{i}^{(j)}_{r}}=1}^{N}F_{{{i}^{(j)}_{r}}}(p).  
\end{align}

The optimization problem is formulated as:
\begin{subequations}
  \begin{alignat}{2}
    &\!\min_{p\in \mathbb{R}^5 }     &\qquad& \mathrm{GCF}(p) = \sum_{{{i}^{(j)}_{r}}=1}^{N}F_{{{i}^{(j)}_{r}}}(p) \label{eqn:saadi_optimisation} \\
    &\text{subject to} &      & \mathrm{LB}\leq p \leq \mathrm{UB}.
  \end{alignat}
\end{subequations}
where $\mathrm{LB}$ and $\mathrm{UB}$  are defined as above.

\FK{Each summand $F_{{{i}^{(j)}_{r}}}(p)$ of equation \ref{eqn:saadi_optimisation},
that represent the cost functionals for individual blocks, 
can be treated independently and thus, offers potential for parallel execution. 
In order to take advantage of parallel execution
we used the R-package \texttt{parallel} to send jobs
to the available processing cores on the computing grid.}

A bootstrap parameter set of the L\"ammer model is identified by running the optimization procedure on two Intel(R) Xeon(R) Gold 6138 processors providing a processor base 
frequency of 2.00GHz with 40 threads each and a total available memory of 502GB
takes three days and 11 hours. Trial values are used as start input parameters for the first bootstrapping optimization procedure. After that the initial values are taken from the first optimization result. 

%%%%%%%%%%%%%%%%%%%%%%%%%%%%%%%%%%%%%%%%%%%%%%%%%%%%%%%%%%%%%%%%%%%%%%%%%%%%%%%%
% Author: Mohamed
\section{Uncertainty Quantification and Parameter Distributions for the L\"{a}mmer Model}
\label{sec:saadi_results}
%%%%%%%%%%%%%%%%%%%%%%%%%%%%%%%%%%%%%%%%%%%%%%%%%%%%%%%%%%%%%%%%%%%%%%%%%%%%%%%%

Before we present the results of the bootstrap method in detail, we briefly explain the set up of our numerical experiments. 

Lifetime models, after calibration with data, should provide accurate lifetime predictions
and generalize for different loading scenarios and different dwell times. To test the model generalization capabilities, we split data into experiments with and without dwell time. The corresponding results can be found in section
\ref{sec:saadi_param-estim-mater}
and 
\ref{sec:saadi_param-fitt-data}, respectively.

%%%%%%%%%%%%%%%%%%%%%%%%%%%%%%%%%%%%%%%%
\subsection{Parameter estimation of materials data with and without dwell time}
\label{sec:saadi_param-estim-mater}
%%%%%%%%%%%%%%%%%%%%%%%%%%%%%%%%%%%%%%%%

In this section, we quantify the uncertainty in parameter identification by the bootstrapping method. The model simulation of Lämmer's damage evolution equation is
compared with the experimentally observed damage based on the low-cycle fatigue datasets BDE6d1 -- BDE6d4 and BDE6dh1 -- BDE6dh3. The identification procedure is based only on data with dwell time. We plot the corresponding damage curve for each parameter set
obtained by bootstrapping.\\ 

%%%%%%%%%%%%%%%%%%%%%%%%%%%%%%%%%%%%%%%%
\subsubsection{Damage by data without dwell time}
\label{sec:saadi_damage-data-without}
%%%%%%%%%%%%%%%%%%%%%%%%%%%%%%%%%%%%%%%%

\begin{figure*}[h]
     \centering
     \begin{subfigure}[b]{0.48\textwidth}
         \centering
         \includegraphics[width=\textwidth]{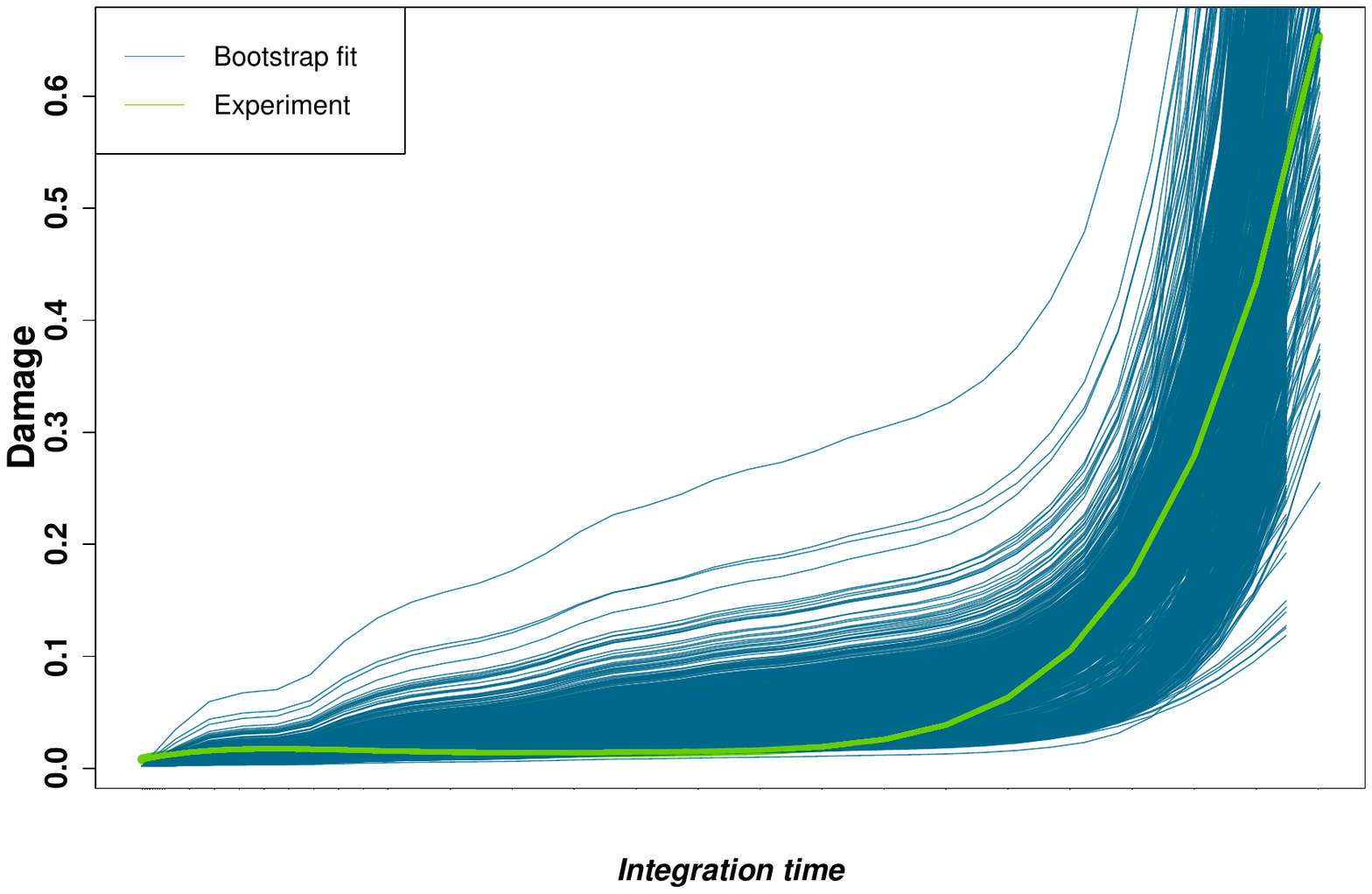}
         \caption{Damage curves by bootstrapping (BDE6d1).}
         \label{fig:saadi_CI_BDE6d1}
     \end{subfigure}
     \hfill
     \begin{subfigure}[b]{0.48\textwidth}
         \centering
         \includegraphics[width=\textwidth]{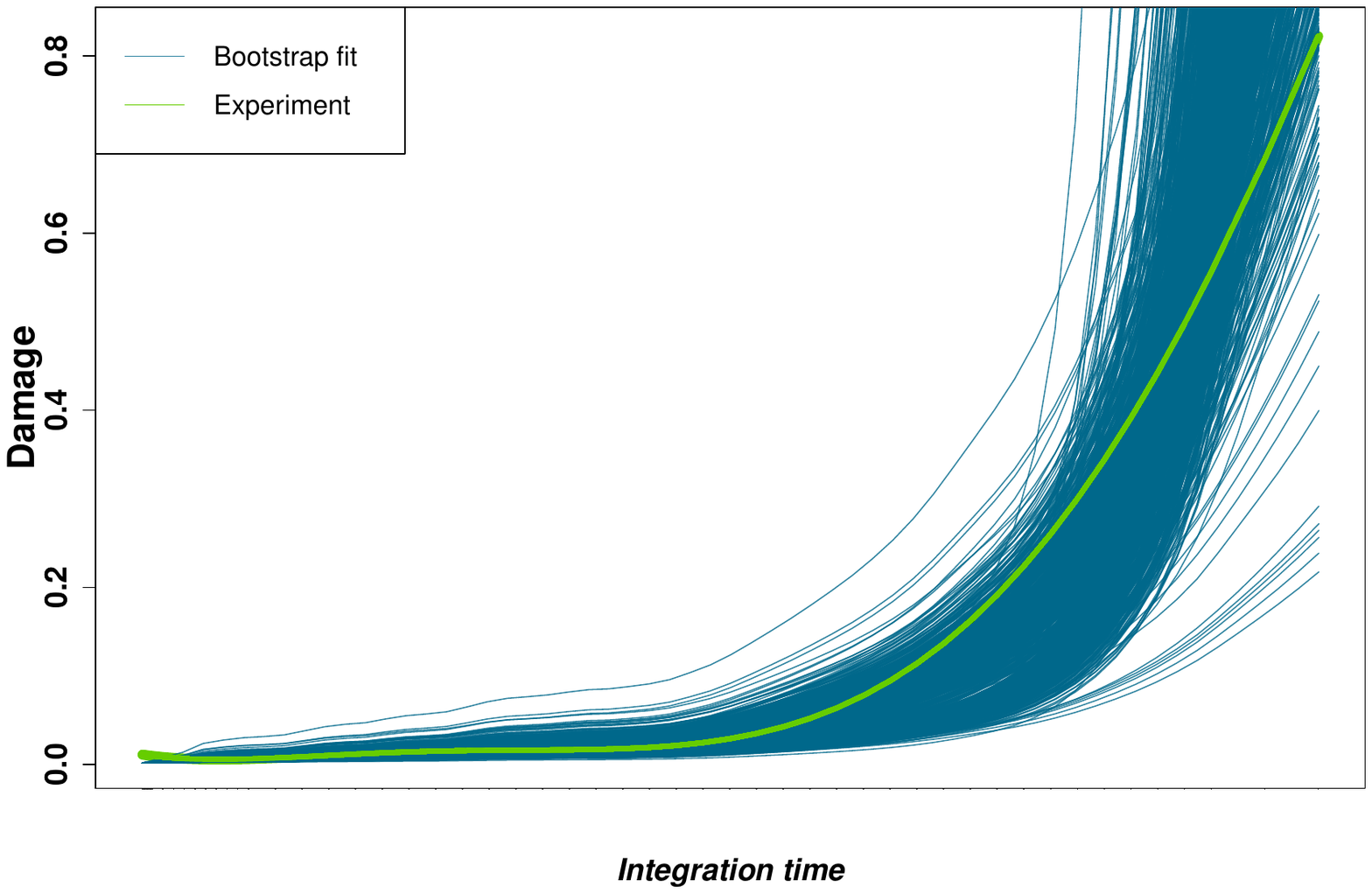}
         \caption{Damage curves by bootstrapping (BDE6d2).}
         \label{fig:saadi_CI_BDE6d2}
     \end{subfigure}
     \vfill
     \begin{subfigure}[b]{0.48\textwidth}
         \centering
         \includegraphics[width=\textwidth]{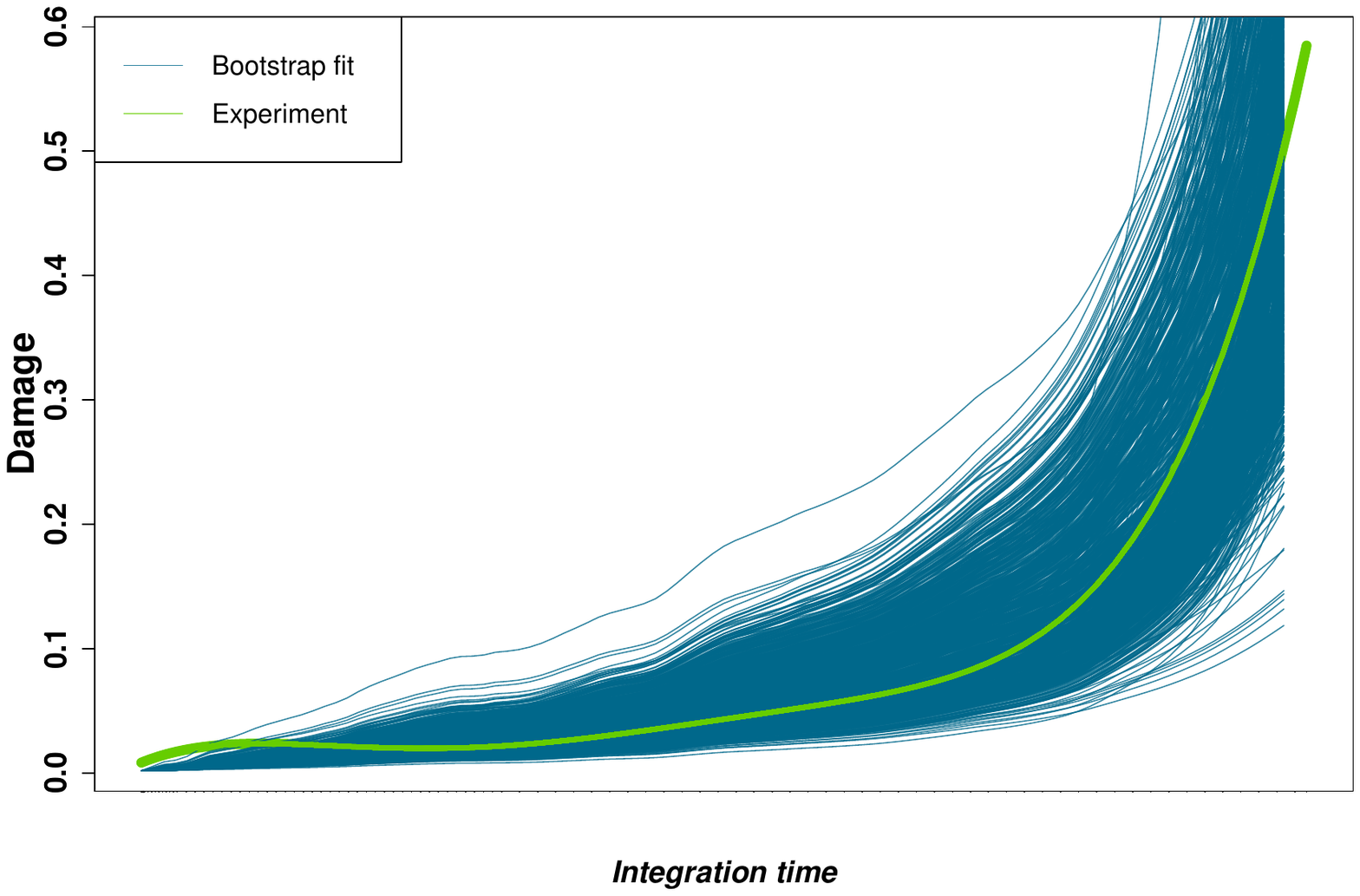}
         \caption{Damage curves by bootstrapping (BDE6d3).}
         \label{fig:saadi_CI_BDE6d3}
     \end{subfigure}
     \hfill
     \begin{subfigure}[b]{0.48\textwidth}
         \centering
         \includegraphics[width=\textwidth]{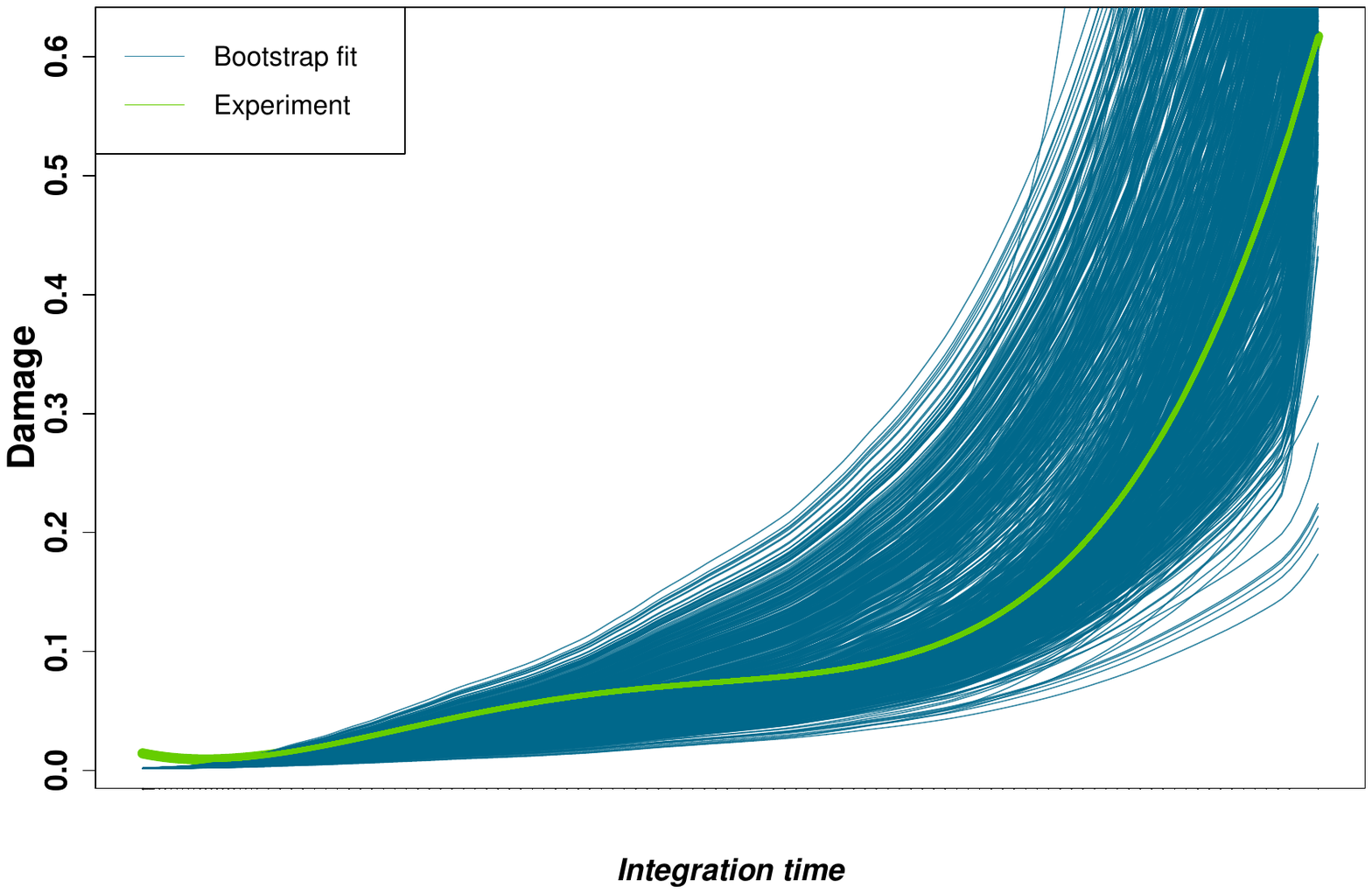}
         \caption{Damage curves by bootstrapping (BDE6d4).}
         \label{fig:saadi_CI_BDE6d4}
     \end{subfigure}
     \caption{Comparison of the estimated damage curves (blue) by 1000 bootstrap simulations and
       the experimental damage curve (green) for the data (BDE6d1-4) without dwell times.}
        \label{fig:saadi_CI_BDE6d_1234}
\end{figure*}

The result of the bootstrapping procedure described in the section \ref{sec:bootstrap} obtained from $1000$ 
bootstrap simulations based on the experimental data without dwell time (BDE6d1 -- BDE6d4)
is shown in figure \ref{fig:saadi_CI_BDE6d_1234}. The green
plots show the experimental damage evolution 
determined as described in section \ref{sec:kzo_data_and_damage_assessment},
and the blue plots show the computed damage evolution based
on equation \ref{eq:kzo_laemmer_damage}
and the bootstrapped parameters $p=(\alpha_0,\alpha_1,\alpha_2,q_0,q_1)$.
 It is observed that the bootstrapped curves envelope the experimental damage over time, i.e. over number of cycles and thereby provide a reasonable quantification of epistemic uncertainty. 

% paragraph taken from discussion
Here the identical bootstrap parameter set is used to predict damage evolution for all 
four experiments, as shown in figure \ref{fig:saadi_CI_BDE6d_1234}.
These results strongly support the claim that it is indeed possible 
to represent the empirically observed damage evolution
by using \eqref{eq:kzo_laemmer_damage} and 
confirms the applicability of the damage evolution equation
to low-cycle fatigue,
which was an open question, see \cite[p. 26]{samir2006_konsitutiveWerkstoffbeschreibungKriechermuedung}.

\begin{figure*}[h]
     \centering
     \begin{subfigure}[b]{0.42\textwidth}
         \centering
         \includegraphics[width=\textwidth]{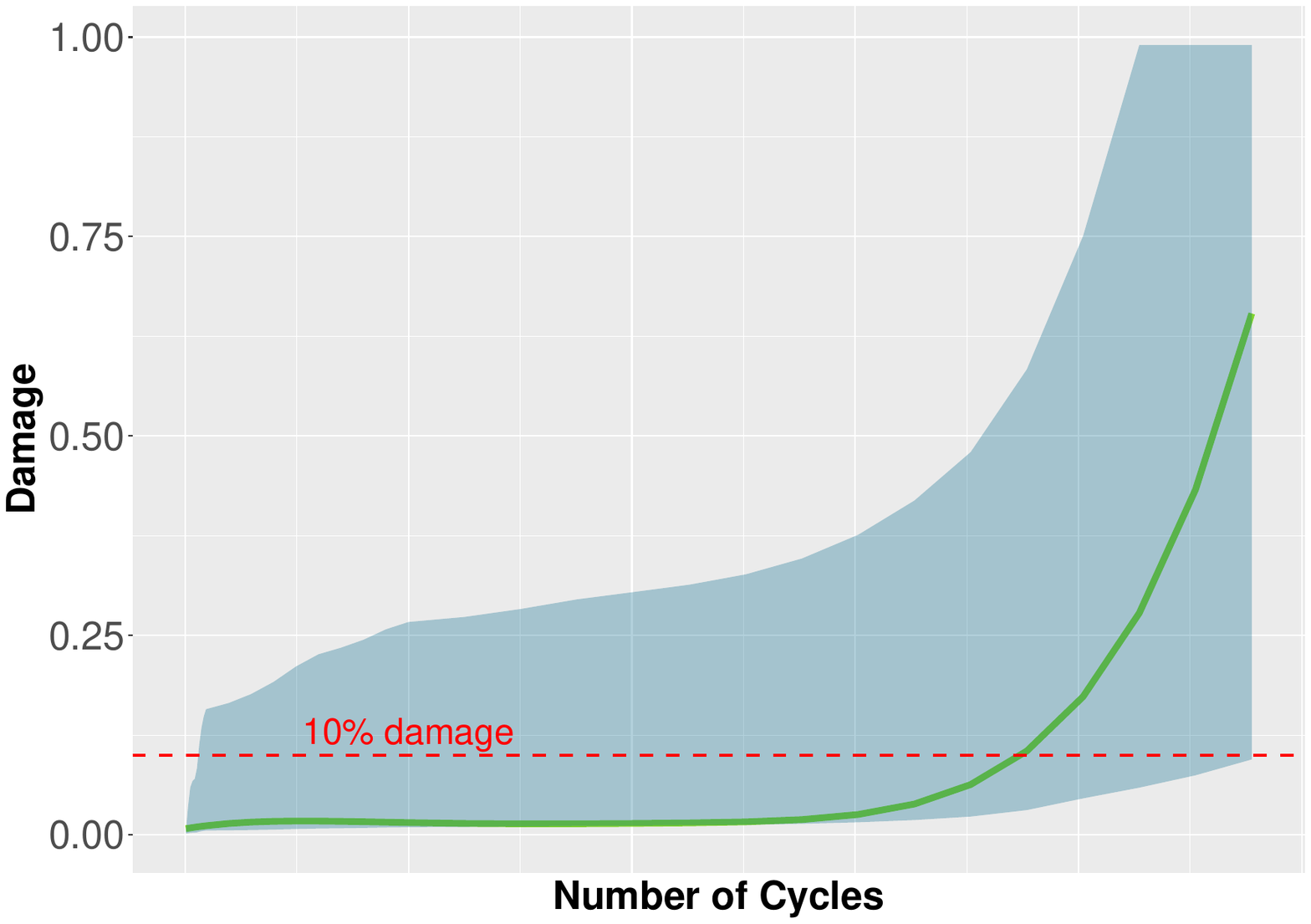}
         \caption{Damage curves with confidence band (BDE6d1).}           
         \label{fig:saadi_bootstrapping_KI_1}
     \end{subfigure}
     \hfill
     \begin{subfigure}[b]{0.42\textwidth}
         \centering
         \includegraphics[width=\textwidth]{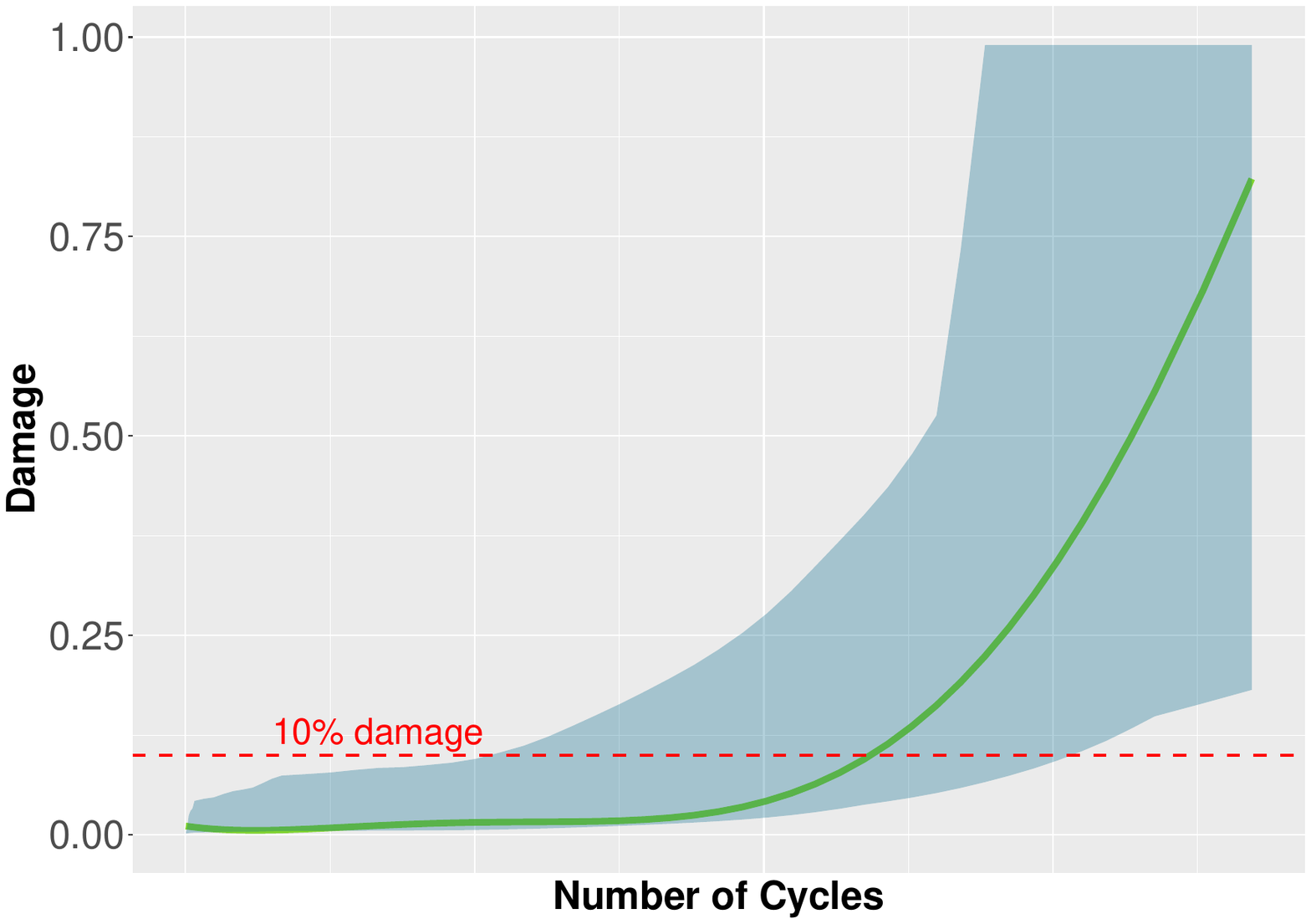}
    		 \caption{Damage curves with confidence band (BDE6d2).}
         \label{fig:saadi_bootstrapping_KI_2}            
     \end{subfigure}
     \vfill
     \begin{subfigure}[b]{0.42\textwidth}
         \centering
		\includegraphics[width=\textwidth]{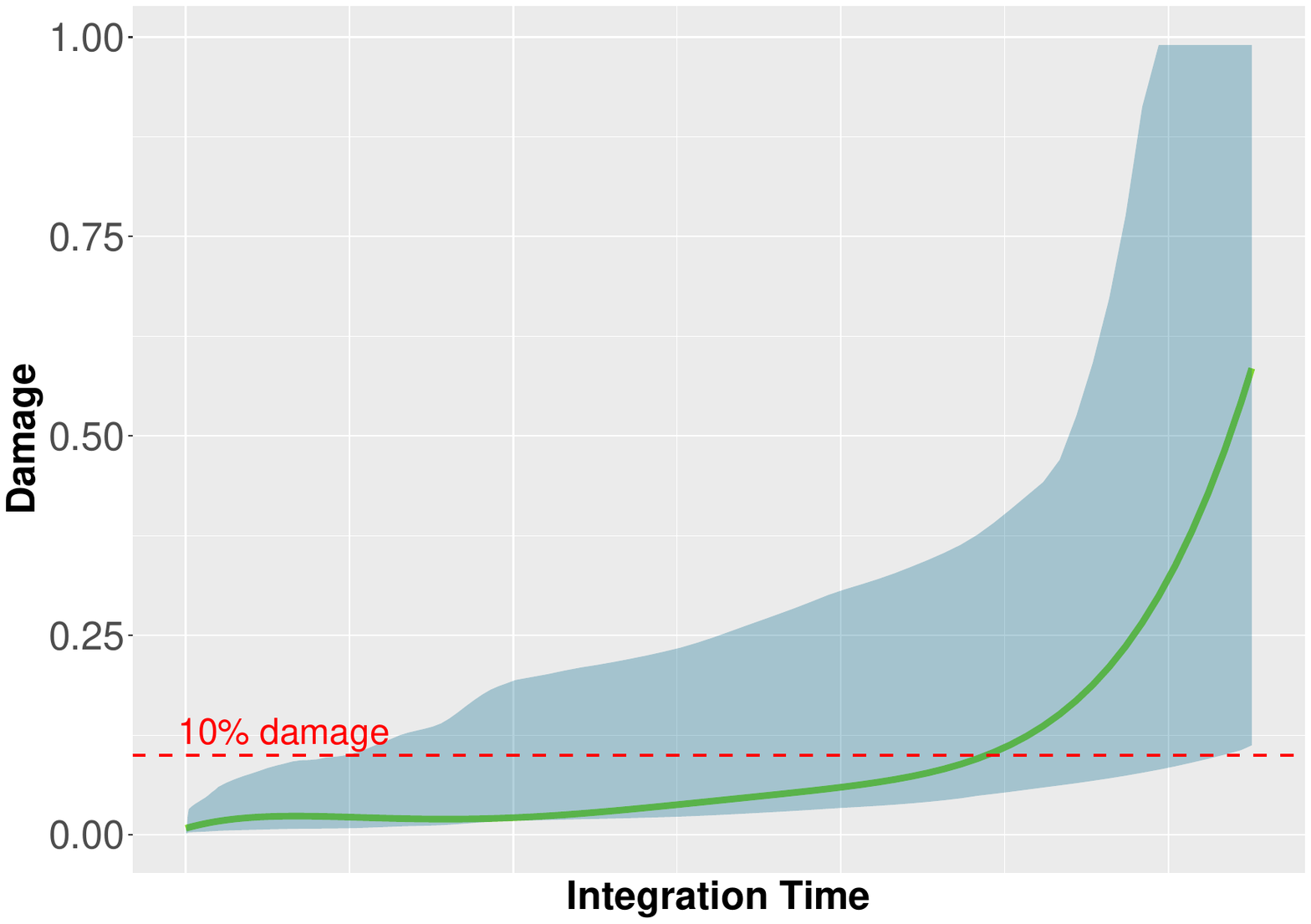}
    		\caption{Damage curves with confidence band (BDE6d3).}
    		\label{fig:saadi_bootstrapping_KI_3}            
     \end{subfigure}
     \hfill
     \begin{subfigure}[b]{0.42\textwidth}
         \centering         
		\includegraphics[width=\textwidth]{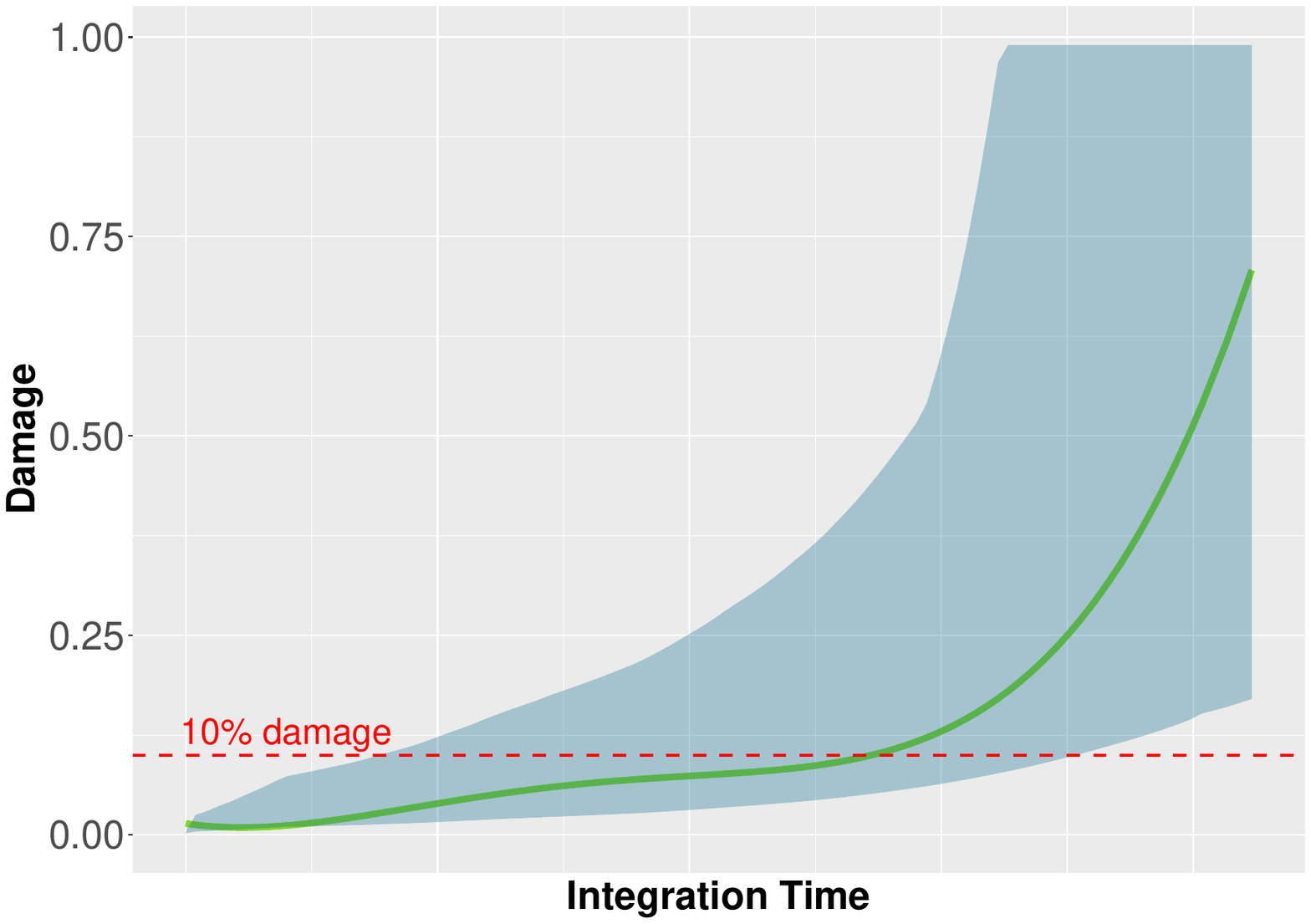}
    		\caption{ Damage curves with confidence band (BDE6d4).}
    		\label{fig:saadi_bootstrapping_KI_4}                
     \end{subfigure}
     \caption{$95\%$ confidence bands of the damage curves based on 1000
     bootstrap simulations. The red horizontal line describes the critical damage parameter $D_{\textrm{crit.}}=0.1$. 
     }
     \label{fig:saadi_bootstrapping_KI_1234}
\end{figure*}

In order to quantify the uncertainties from the bootstrap process, the experimental
damage curve with $95\%$ confidence bands are plotted in figure \ref{fig:saadi_bootstrapping_KI_1234}. It can be
seen that for data without dwell time, the confidence bands 
of all bootstrapped curves contain the experimental damage curves.

% paragraph taken from discussion
So far, there is no indication that the model parameters depend on the strain range of the performed
experiment. This
is in contrast to the findings presented by \cite{khan2018_damageEvolutionOfHighChromiumSteels},
who considered the damage evolution equation suggested by  
\cite{lemaitreChaboche2009_mechanicsOfSolidMaterials} and reported a model parameter dependency on the strain range that is applied.

%%%%%%%%%%%%%%%%%%%%%%%%%%%%%%%%%%%%%%%%
\subsubsection{Block-Size Independence}
\label{sec:block_size_independence}
%%%%%%%%%%%%%%%%%%%%%%%%%%%%%%%%%%%%%%%%

Since the choice of the block length of the applied bootstrap procedure is somehow arbitrary, the influence of different block lengths has to be studied. Therefore, we restart our bootstrapping algorithm using four different block length 
that contain $(275,325,425,475)$ number of cycles.
For each size we plot the damage curves of the $100$ bootstrap sets and the 
the experimentally observed damage evolution.
The results of this ablation study are given in appendix \ref{sec:appendix_A1}.

%%%%%%%%%%%%%%%%%%%%%%%%%%%%%%%%%%%%%%%%
\subsubsection{Distributions of number of cycles and model parameters}
\label{sec:saadi_distr-numb-cycl}
%%%%%%%%%%%%%%%%%%%%%%%%%%%%%%%%%%%%%%%%

\begin{figure*}[t]
     \centering
     \begin{subfigure}[b]{0.40\textwidth}
         \centering
         \includegraphics[width=\textwidth]{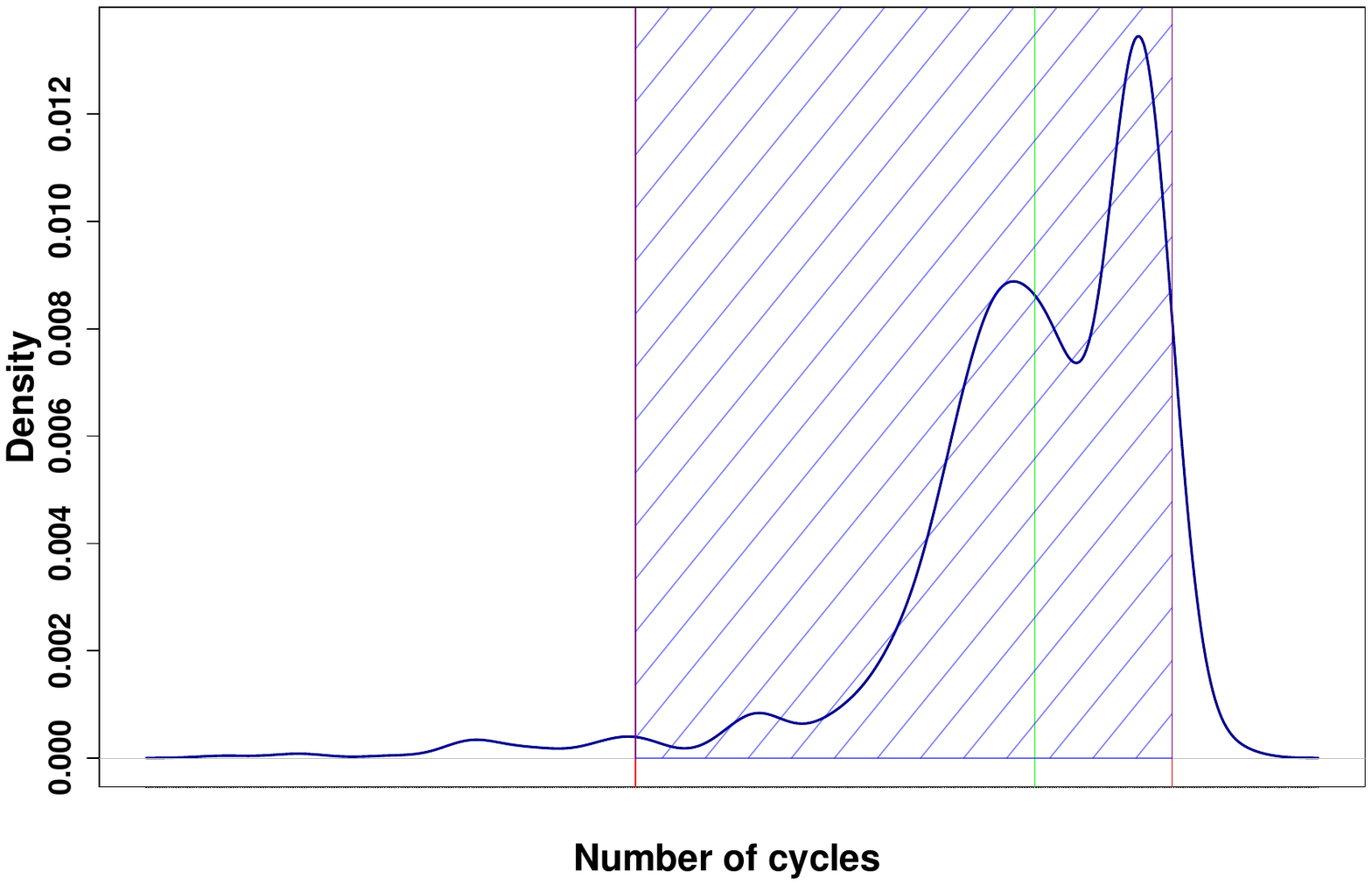}
         \caption{CI of number of cycles to crack initiation (BDE6d1).}
         \label{fig:saadi_CI_BDE6d1_time}
     \end{subfigure}
     \hfill
     \begin{subfigure}[b]{0.40\textwidth}
         \centering
         \includegraphics[width=\textwidth]{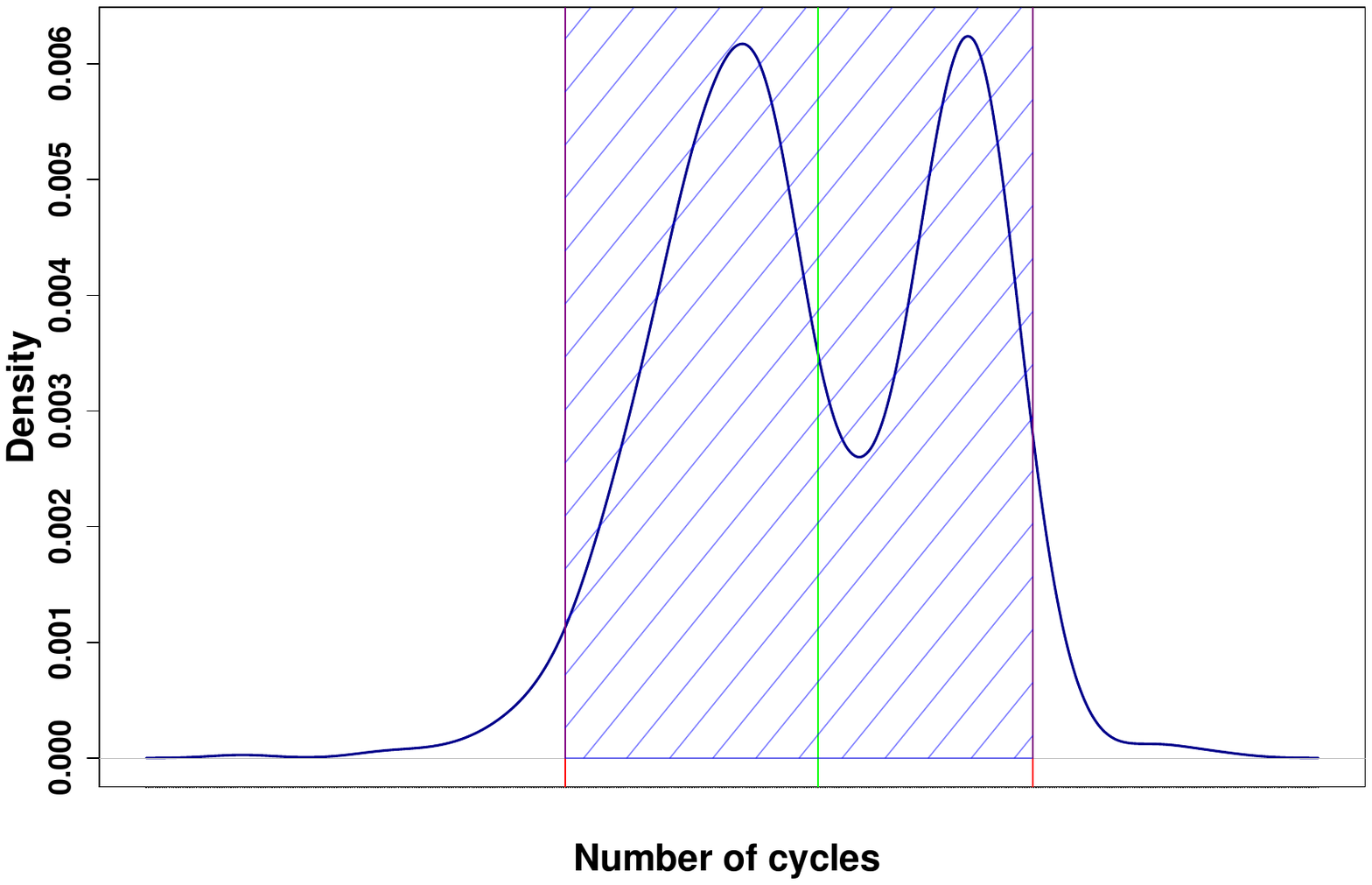}
         \caption{CI of number of cycles to crack initiation (BDE6d2).}
         \label{fig:saadi_CI_BDE6d2_time}
     \end{subfigure}
     \vfill
     \begin{subfigure}[b]{0.40\textwidth}
         \centering
         \includegraphics[width=\textwidth]{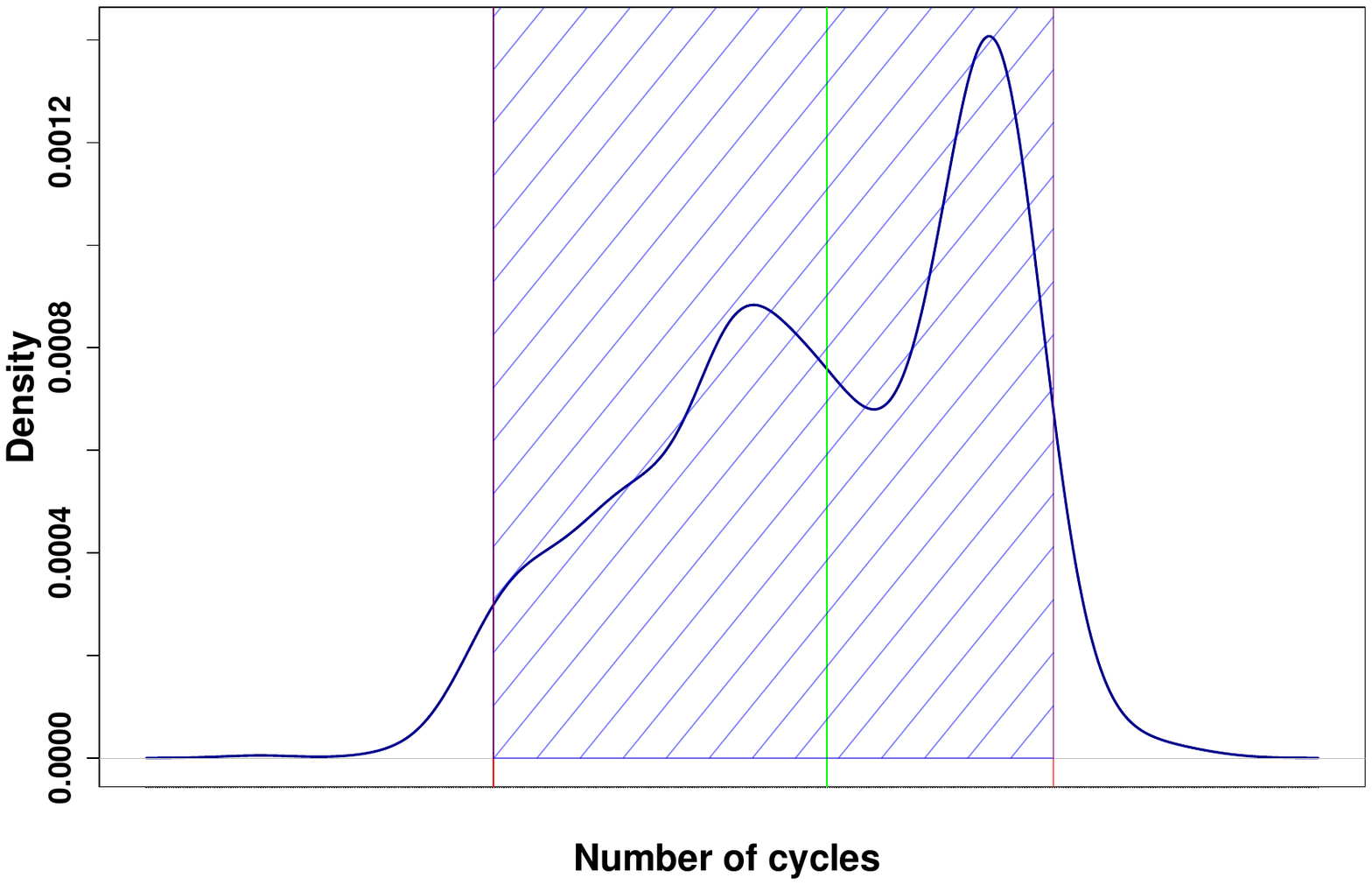}
         \caption{CI of number of cycles to crack initiation (BDE6d3).}
         \label{fig:saadi_CI_BDE6d3_time}
     \end{subfigure}
     \hfill
     \begin{subfigure}[b]{0.40\textwidth}
         \centering
         \includegraphics[width=\textwidth]{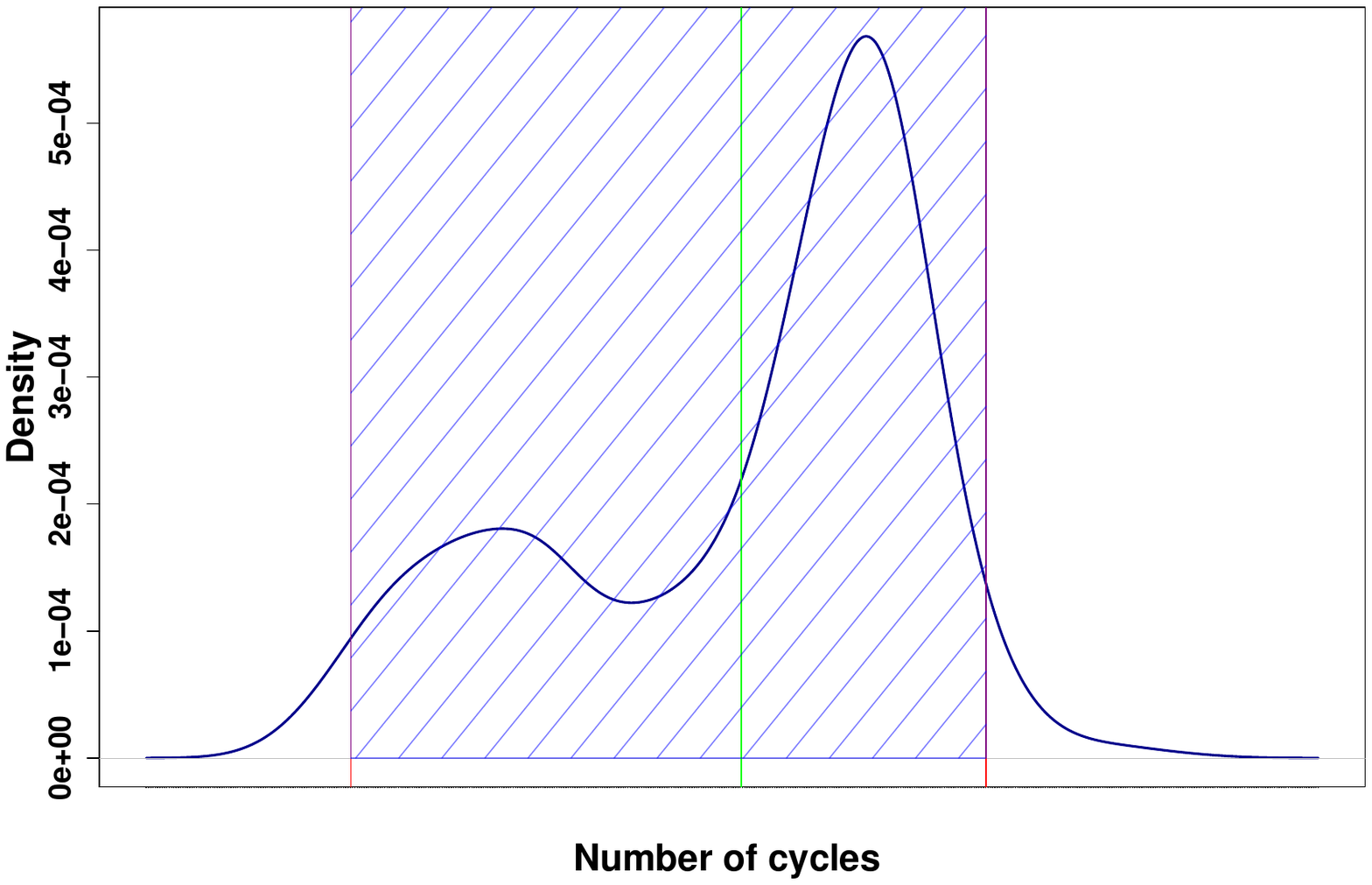}
         \caption{CI of number of cycles to crack initiation (BDE6d4).}
         \label{fig:saadi_CI_BDE6d4_time}
     \end{subfigure}
     \caption{$95\%$ confidence intervals of number of cycles to crack initiation (vertical green line shows the mean).}
        \label{fig:saadi_CI_BDE6d1234_time}
\end{figure*}

The distributions of the number of cycles to crack initiation
for the four datasets without dwell time is shown in figure \ref{fig:saadi_CI_BDE6d1234_time}. 
Areas between the red lines show the $95\%$ confidence intervals for the mean values (green lines). 

\begin{figure*}[t]
     \centering
     \begin{subfigure}[b]{0.40\textwidth}
         \centering
         \includegraphics[width=\textwidth]{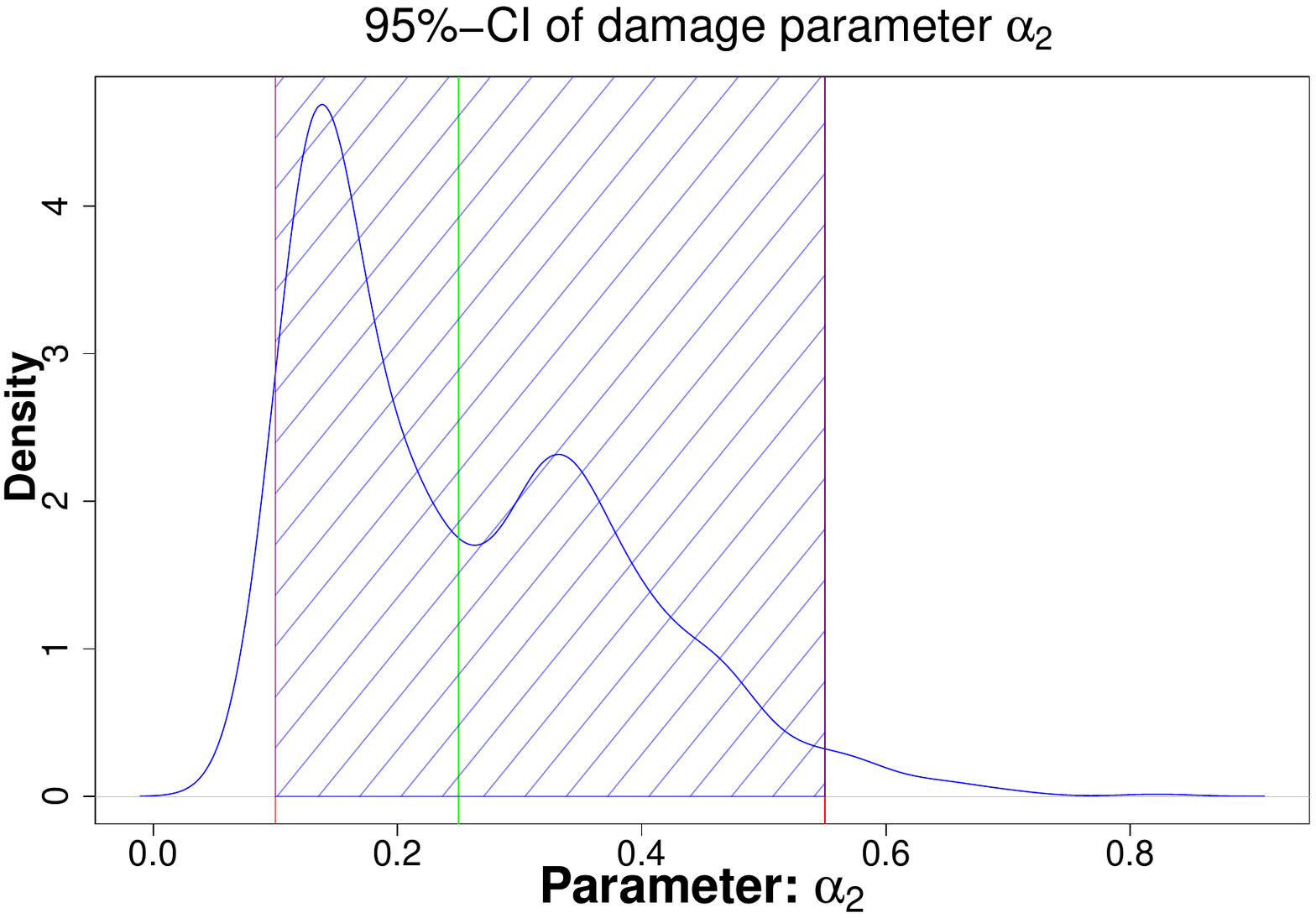}
         \caption{CI of parameter $\alpha_2$.}
         \label{fig:saadi_CI_BDE6d2_param}
     \end{subfigure}
     \vfill
     \begin{subfigure}[b]{0.40\textwidth}
         \centering
         \includegraphics[width=\textwidth]{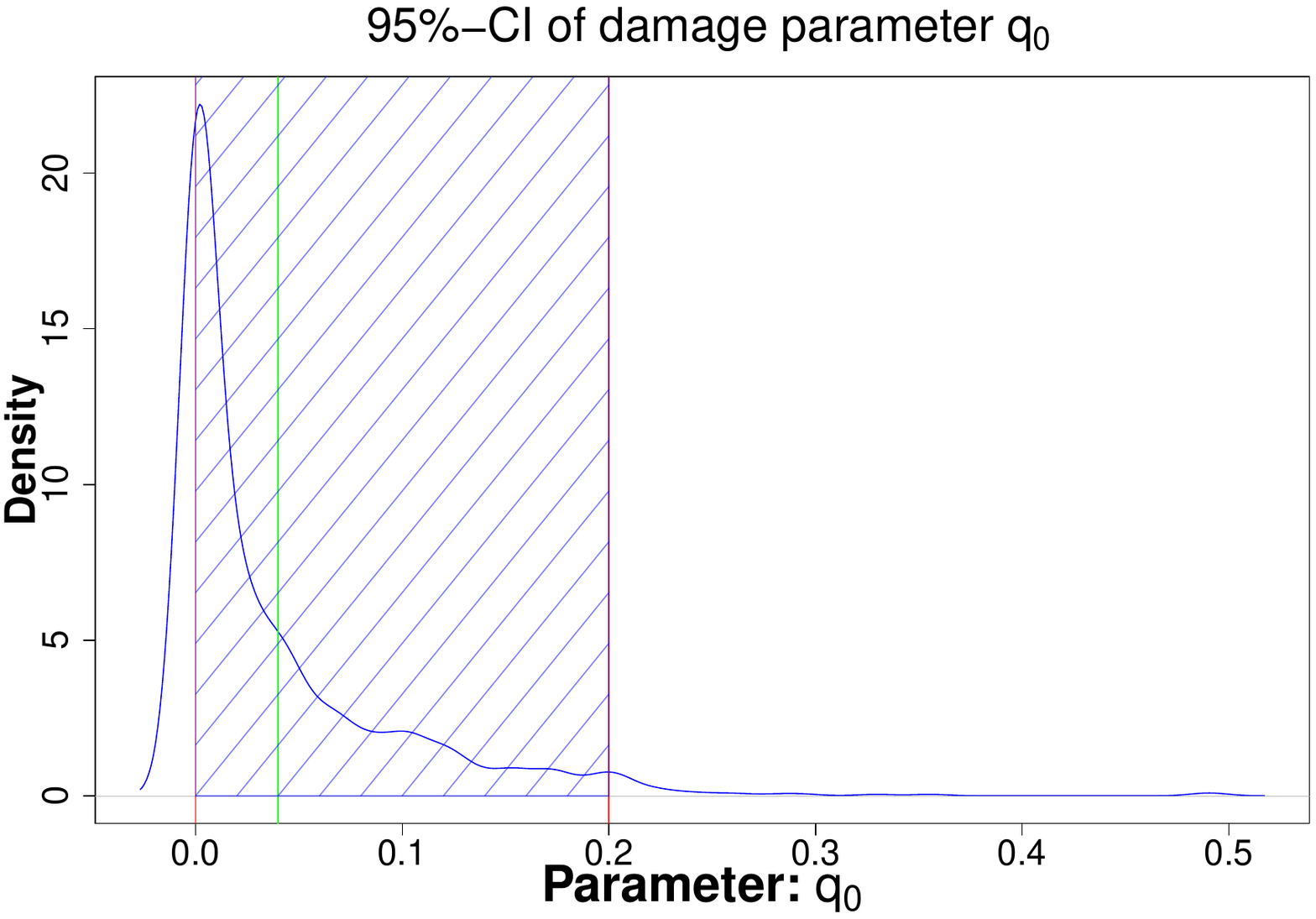}
         \caption{CI of parameter $q_0$.}
         \label{fig:saadi_CI_BDE6d3_param}
     \end{subfigure}
     \hfill
     \begin{subfigure}[b]{0.40\textwidth}
         \centering
         \includegraphics[width=\textwidth]{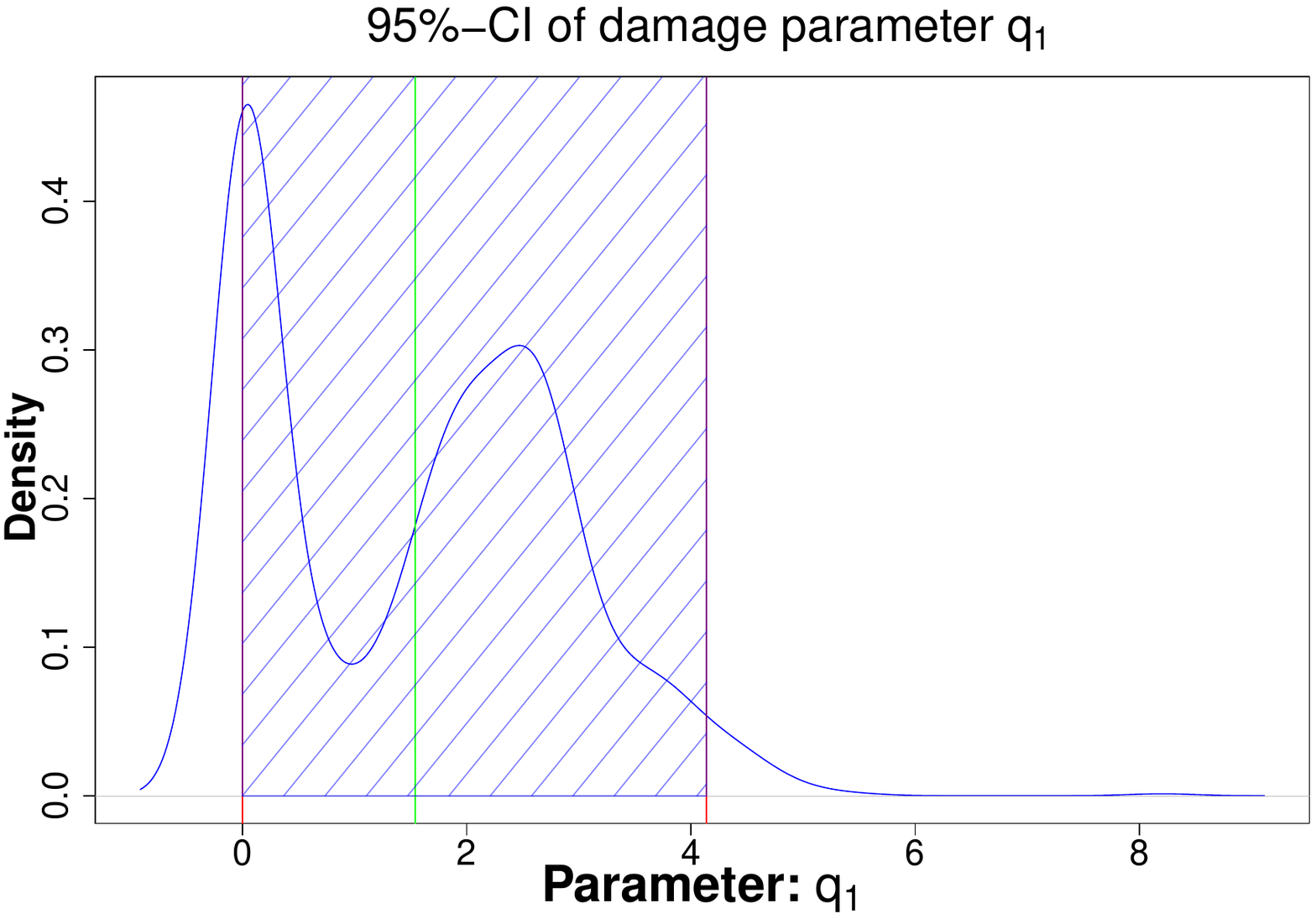}
         \caption{CI of parameter $q_1$.}
         \label{fig:saadi_CI_BDE6d4_param}
     \end{subfigure}
     \caption{$95\%$ Confidence intervals of material parameters $\alpha_2, q_0$ and $q_1$.}
        \label{fig:saadi_CI_BDE6d1234_param}
\end{figure*}

% Parameters distributions
The distributions of the 
damage model parameters $\alpha_2, q_0$ and $q_1$ that belongs to the
damage evolution equation proposed by Lämmer (\ref{eq:kzo_laemmer_damage})
is shown in figure \ref{fig:saadi_CI_BDE6d1234_param}.
The vertical red lines show the
ranges of $95\%$ confidence intervals, and the
vertical green lines show the damage parameters' means.

% paragraph taken from discussion
The parameters $\alpha_0$ and $\alpha_1$  are distributed  symmetrically around zero and, 
therefore, are  as non-significant in a statistical sense. We thus omitted them in figure \ref{fig:saadi_CI_BDE6d1234_param}.
Especially, $\alpha_0=0$ can be
understood because the material investigated, as mentioned in section \ref{sec:kzo_data_set_origin}
is high-chromium cast steel, including
internal defects due to the manufacturing process. These internal defects 
are within tolerable limits and, thus, are common to this kind of materials, see \cite[fig. 5.8, p. 35]{Jobski2018_AVIFVorhaben307_finalreport}.
These internal defects are expected to grow under creep-fatigue loading and dominate
the end-of-life, which is modeled
by the second term of equation \eqref{eq:kzo_laemmer_damage}.
Unfortunately, a explanation of why the model parameter $\alpha_1$ appears to be non-significant cannot be given at the current state of research.

From a statistical perspective, the parameter distributions in Figure \ref{fig:saadi_CI_BDE6d1234_param} of $\alpha_2$ and $q_1$ show multimodal behavior. This shows that in the context of CDM, different parameter combinations lead to similar fitting curves, which makes it difficult to disambiguate the parameter values and consider them as material properties that are measurable in the given experiments. 

Furthermore, as a result of the determined bootstrap parameters sets, a negative correlation coefficient 
$\hat \rho_{\alpha_2,q_1}=-0.671$ can be computed between $\alpha_2$ and $q_1$.  Such correlations 
are usually detected via the bootstrapping approach and show that an uncertainty quantification 
by parameter does not properly represent the actual model uncertainty. The parameter 
$q_0$ is only weakly correlated to $\alpha_2$
{( $\hat \rho_{q_0,\alpha_2}=0.209$)} and $q_1$ ($\hat \rho_{q_0,q_1}=-0.079$).

%%%%%%%%%%%%%%%%%%%%%%%%%%%%%%%%%%%%%%%%
\subsection{Parameter fitting of data with dwell time}
\label{sec:saadi_param-fitt-data}
%%%%%%%%%%%%%%%%%%%%%%%%%%%%%%%%%%%%%%%%

In this section, we investigate whether or not the damage evolution equation \eqref{eq:kzo_laemmer_damage}
is capable of predicting damage evolution also for creep-fatigue data. Therefore, we fit the Lämmer model parameters for data with dwell times only.

%%%%%%%%%%%%%%%%%%%%%%%%%%%%%%%%%%%%%%%%
\subsubsection{Damage by data with dwell time}
\label{sec:saadi_damage-data-with}
%%%%%%%%%%%%%%%%%%%%%%%%%%%%%%%%%%%%%%%%

\begin{figure*}[t]
\centering
     \begin{subfigure}[b]{0.49\textwidth}
     \centering
         \includegraphics[width=\textwidth]{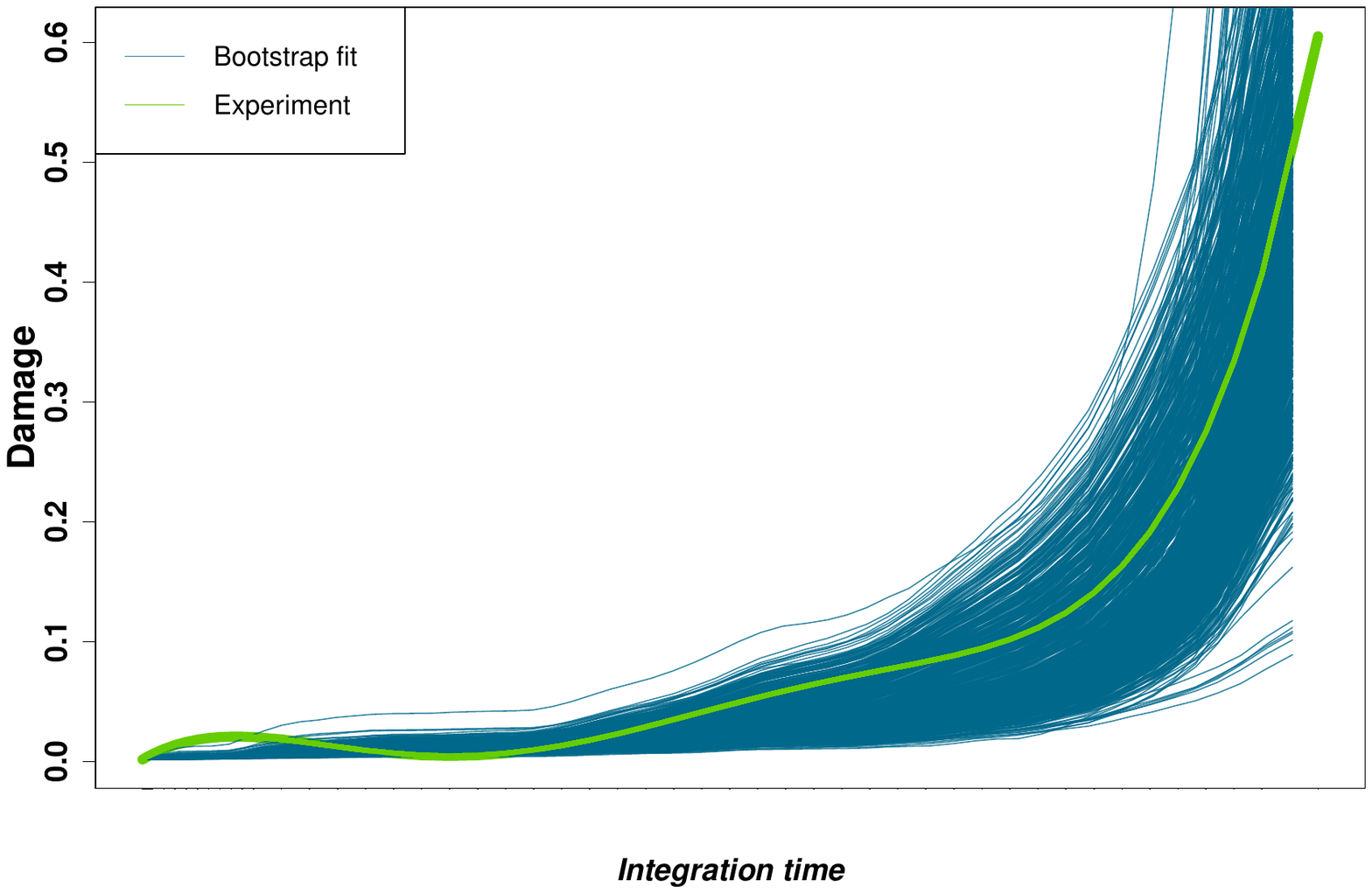}
         \caption{Damage curves bootstrapping (BDE6dh1).}
         \label{fig:saadi_CI_BDE6d1h_only}
     \end{subfigure}
     \vfill
     \begin{subfigure}[b]{0.49\textwidth}
     \centering
         \includegraphics[width=\textwidth]{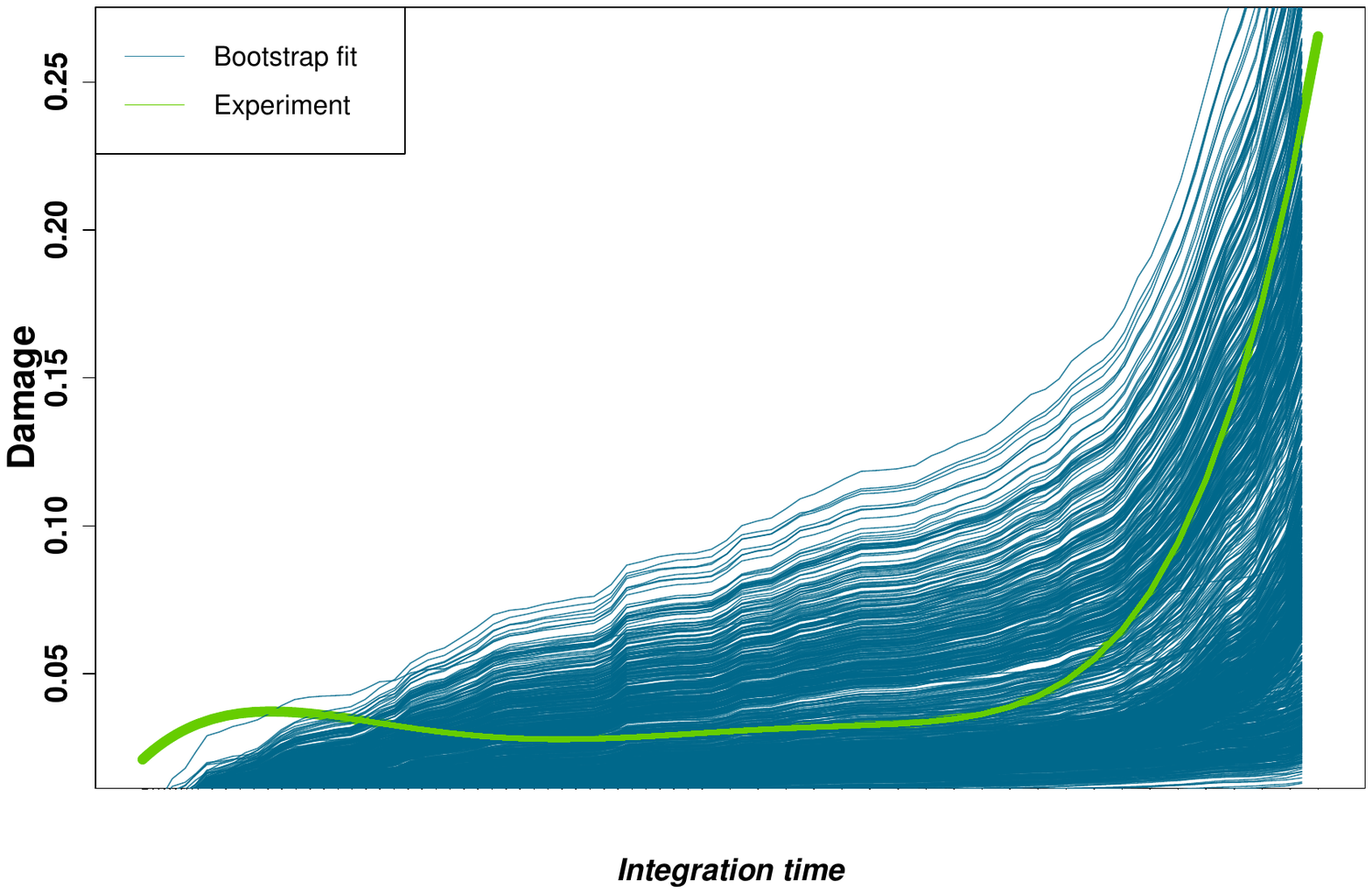}
         \caption{Damage curves by bootstrapping (BDE6dh2).}
         \label{fig:saadi_CI_BDE6d2h_only}
     \end{subfigure}
     \hfill
     \begin{subfigure}[b]{0.49\textwidth}
     \centering
         \includegraphics[width=\textwidth]{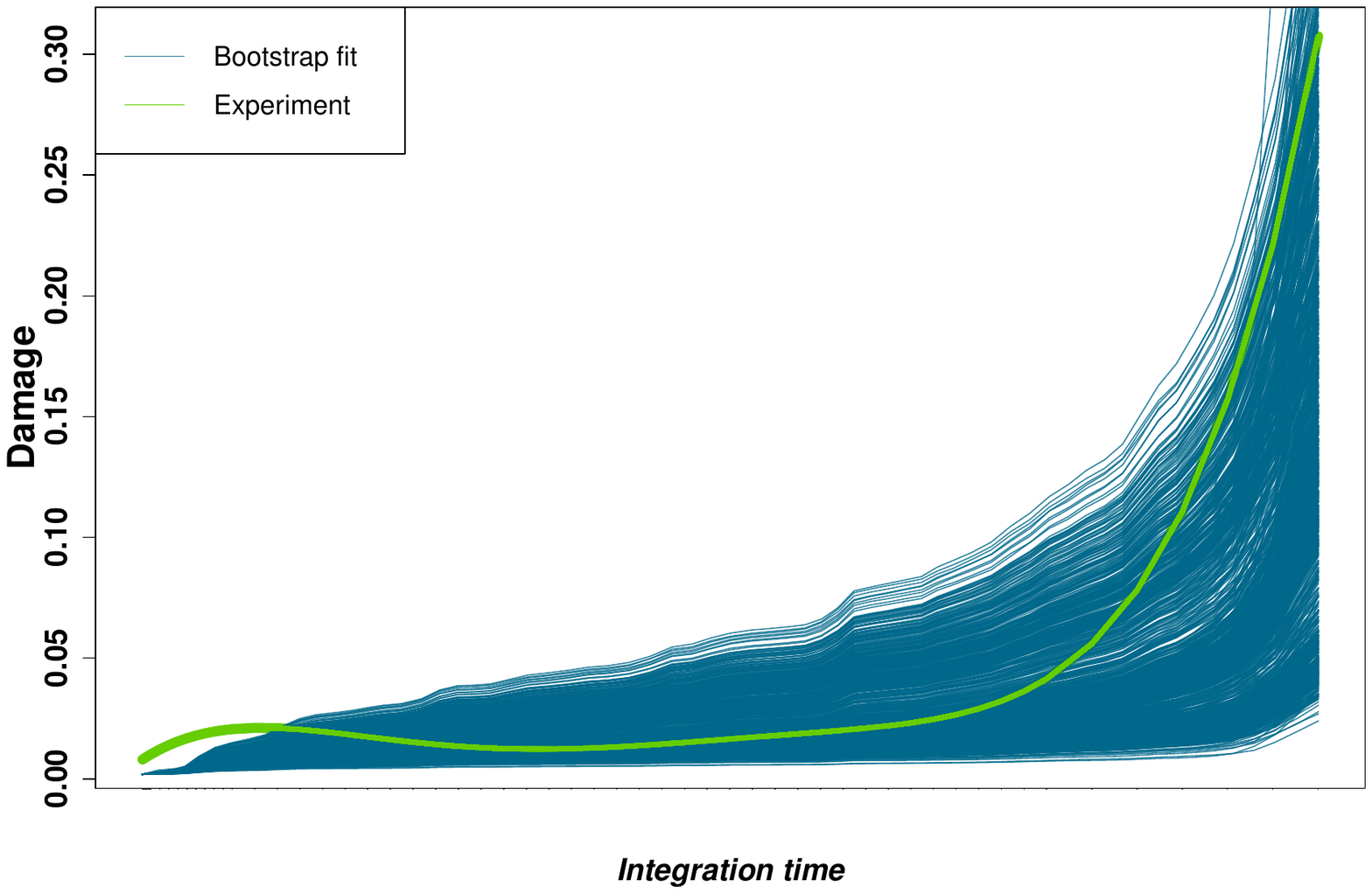}
         \caption{Damage curves by bootstrapping (BDE6dh3).}
         \label{fig:saadi_CI_BDE6d3h_only}
     \end{subfigure}
     \caption{Estimated damage curves (blue) and 1000 bootstrap simulations and the
       experimental damage curve (green) for (BDE6dh\{1,2,3\}) with dwell times.}
        \label{fig:saadi_CI_BD}
        \label{fig:saadi_CI_BDE6dh123_only}
\end{figure*}

The result of the bootstrapping procedure from 1000 simulations considering experimental data with dwell 
times is shown in figure \ref{fig:saadi_CI_BDE6dh123_only}.
Here, experiment BDE6dh3 has a larger dwell time compared to the BDE6dh1 and BDE6dh2, see table 
\ref{tab:kzo_lcf_experiments_overview}.

\begin{figure*}[t]
\centering
     \begin{subfigure}[b]{0.49\textwidth}
     \centering
         \includegraphics[width=\textwidth]{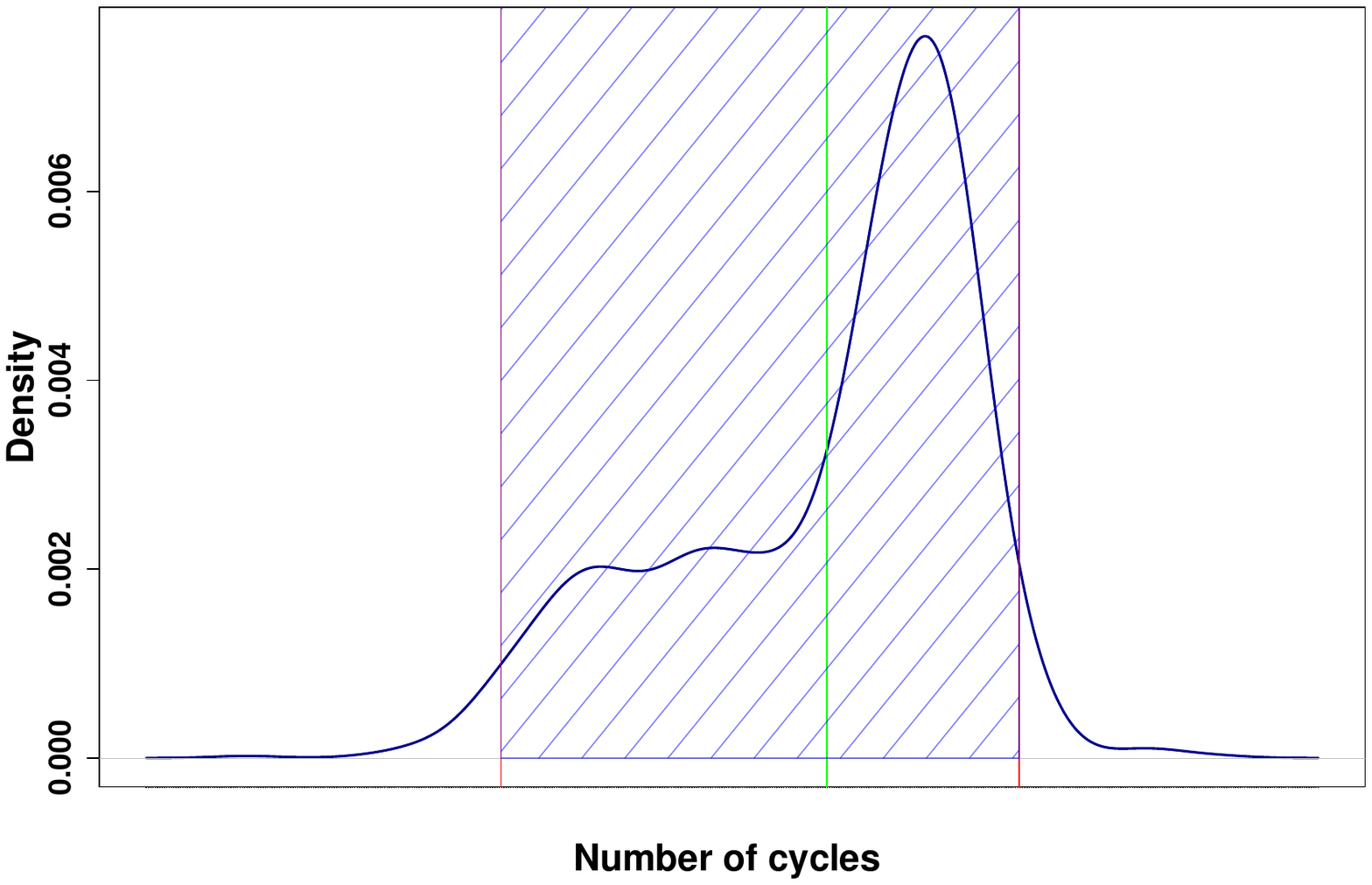}
         \caption{CI of number of cycles to crack initiation (BDE6dh1).}
         \label{fig:saadi_CI_BDE6dh1_time_only}
     \end{subfigure}
     \vfill
     \begin{subfigure}[b]{0.49\textwidth}
     \centering
         \includegraphics[width=\textwidth]{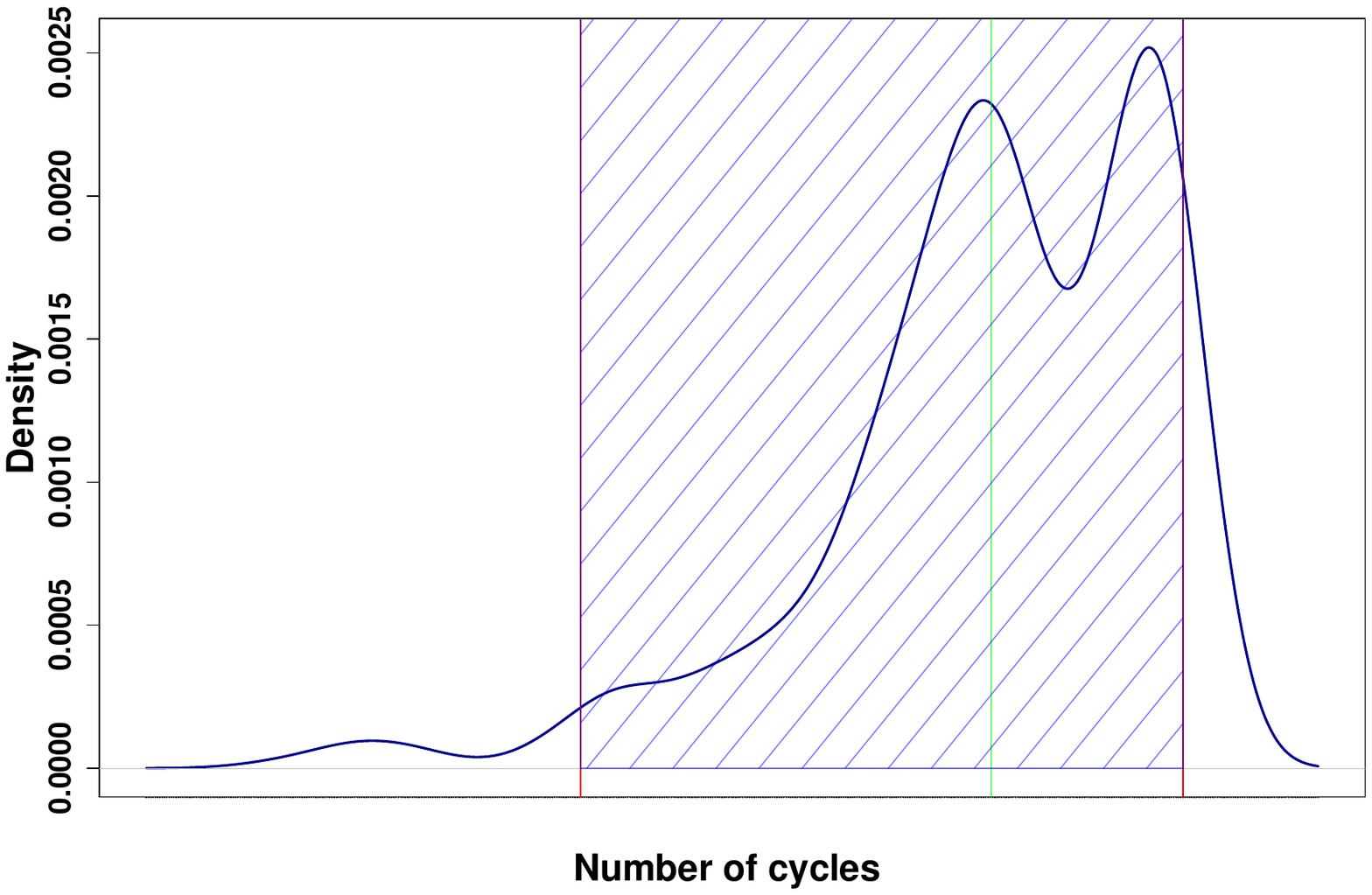}
         \caption{CI of number of cycles to crack initiation (BDE6dh2).}
         \label{fig:saadi_CI_BDE6dh2_time_only}
     \end{subfigure}
     \hfill
     \begin{subfigure}[b]{0.49\textwidth}
     \centering
         \includegraphics[width=\textwidth]{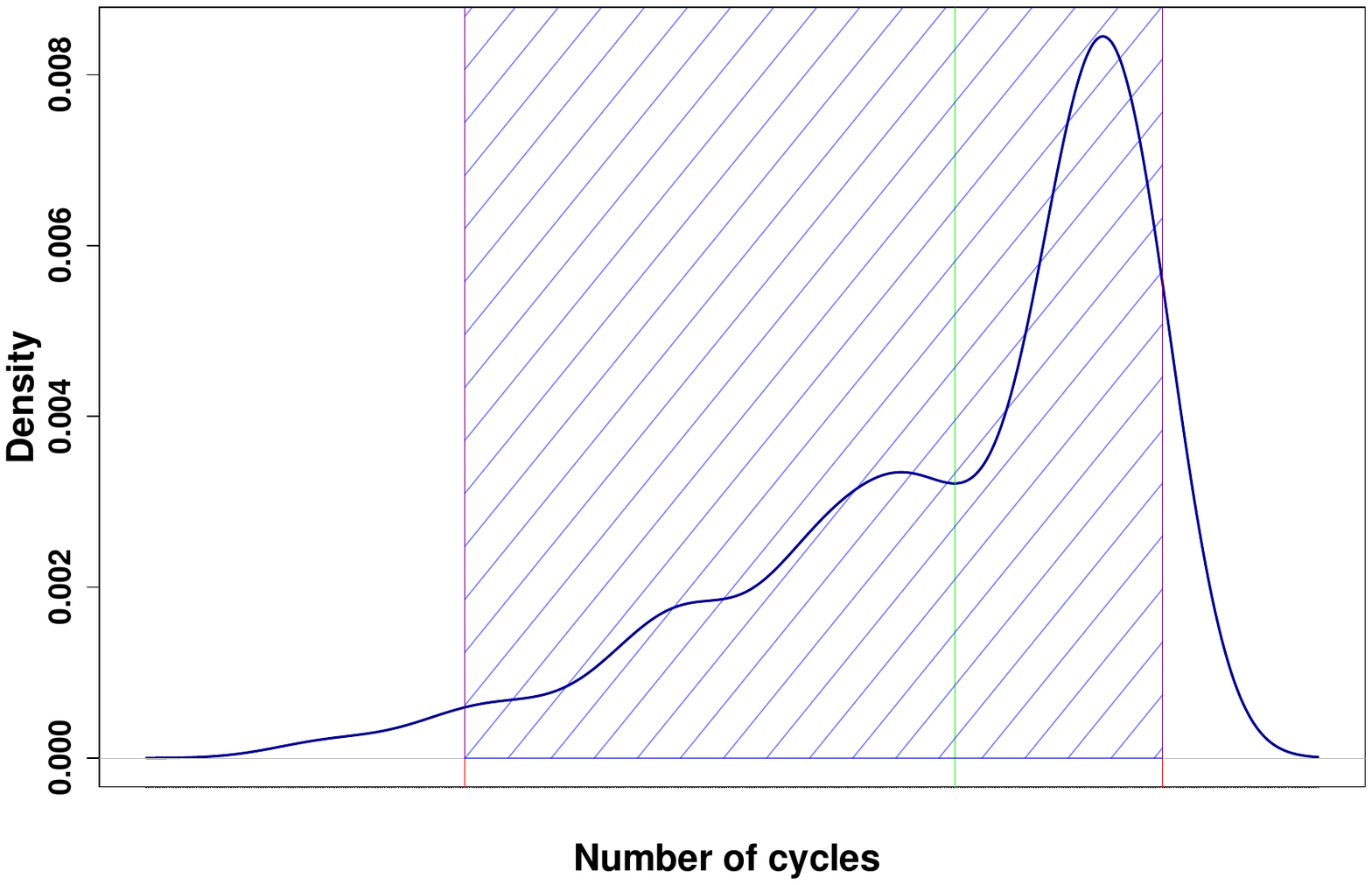}
         \caption{CI of number of cycles  to crack initiation (BDE6dh3).}
         \label{fig:saadi_CI_BDE6dh3_time_only}
     \end{subfigure}
     \caption{$95\%$ confidence intervals of number of cycles to crack initiation (BDE6dh1-3, vertical green line shows the mean).}
     \label{fig:saadi_CI_BDE6dh123_time_only}
\end{figure*} 

The model parameters bootstrap distribution is obtained leaving the estimation procedure unchanged. 

Again, the bootstrap distribution of curves envelopes the observed damage 
evolution and hence, it can be concluded that our
bootstrap based uncertainty quantification is
also applicable for experiments including dwell times. 
 
%%%%%%%%%%%%%%%%%%%%%%%%%%%%%%%%%%%%%%%% 
\subsubsection{Distributions of number of cycles for data with dwell time}
\label{sec:saadi_distr-numb-cycl-1}
%%%%%%%%%%%%%%%%%%%%%%%%%%%%%%%%%%%%%%%%

The distributions of the number of cycles to crack initiation for data with dwell times are
shown in figure \ref{fig:saadi_CI_BDE6dh123_time_only}. Areas between the red
lines show the $95\%$ confidence intervals for the mean values (green lines). The findings on multimodality and correlation of parameters are very similar to those in Section \ref{sec:saadi_distr-numb-cycl} and therefore are not repeated here.

\newpage
\section{Conclusion}
\label{sec:conclusion}
%%%%%%%%%%%%%%%%%%%%%%%%%%%%%%%%%%%%%%%%%%%%%%%%%%%%%%%%%%%%%%%%%%%%%%%%

Common CDM models and their life prediction in most cases do not cover practical 
questions regarding parameter fitting and uncertainty quantification for the predicted lifetime. 

% paragraph
In this paper, we apply the
damage evolution equation \eqref{eq:kzo_laemmer_damage} to low-cycle and creep-fatigue
experiments at \SI{600}{\celsius} and also quantify uncertainty of the predicted damage evolution. To do so, we developed a bootstrap procedure using a sampling over blocks of cycles approach.
The uncertainty within the predicted damage evolution, model parameters and the predicted lifetime were quantified by their respective distributions on the basis of 1000 bootstrap simulations, which have been generated by committing a considerable amount of high-performance computing resources.

As we find, a sophisticated damage evolution equation 
introduced by 
\cite{Laemmer1998_ThermoplastizitatUndThermoviskoplastizitatMitSchadigungBeiKleinenUndGroenDeformationen} comes 
with a considerable epistemic uncertainty and a corresponding spread in the distribution of the predicted 
component life. We furthermore find that some of the model parameters can be eliminated as statistically 
insignificant and multimodal parameter distributions occur, which complicate a direct interpretation of model 
parameters as material properties.

Our contribution is suited to further develop
guidelines and safety factors for lifetime prediction of high-temperature components
using CDM models, which are pretty common
for damage accumulation models
\cite{british_energy_R5_2003, ASME2008_ASMEBoilerandPressureVesselCode, RCCMRx2012_DesignandConstructionRules}

\paragraph*{Acknowledgement.} 

The authors would like to thank the 'Forschungsvereinigung der Arbeitsgemeinschaft der Eisen und Metall verarbeitenden Industrie e.V.' (AVIF No. A316) for their financial support and the working group of the German power plant and gas turbine industry for the very fruitful discussions.

%%%%%%%%%%%%%%%%%%%%%%%%%%%%%%%%%%%%%%%%%%%%%%%%%%%%%%%%%%%%%%%%%%%%%%%%
\pagebreak
%\clearpage
%\appendix
\begin{appendices}
%%%%%%%%%%%%%%%%%%%%%%%%%%%%%%%%%%%%%%%%%%%%%%%%%%%%%%%%%%%%%%%%%%%%%%%%
\section{Comparing bootstrap results for different block length}\label{sec:appendix_A1}

\balance
Finally, the block length of the bootstrapping procedure is documented 
in appendix. As far as determined,
the studied block length (275, 325, 425, 475) does not have a visual impact on
the damage evolution. The study was done for all presented experiments without dwell time, 
but it is only shown for BDE6d2
and BDE6d4 in figure \ref{fig:saadi_bootstrapping_block_size_2} and 
\ref{fig:saadi_bootstrapping_block_size_4}, 
respectively. 

\begin{figure*}[t]
\centering
     \begin{subfigure}[b]{0.49\textwidth}
     \centering
     \includegraphics[width=0.99\textwidth]{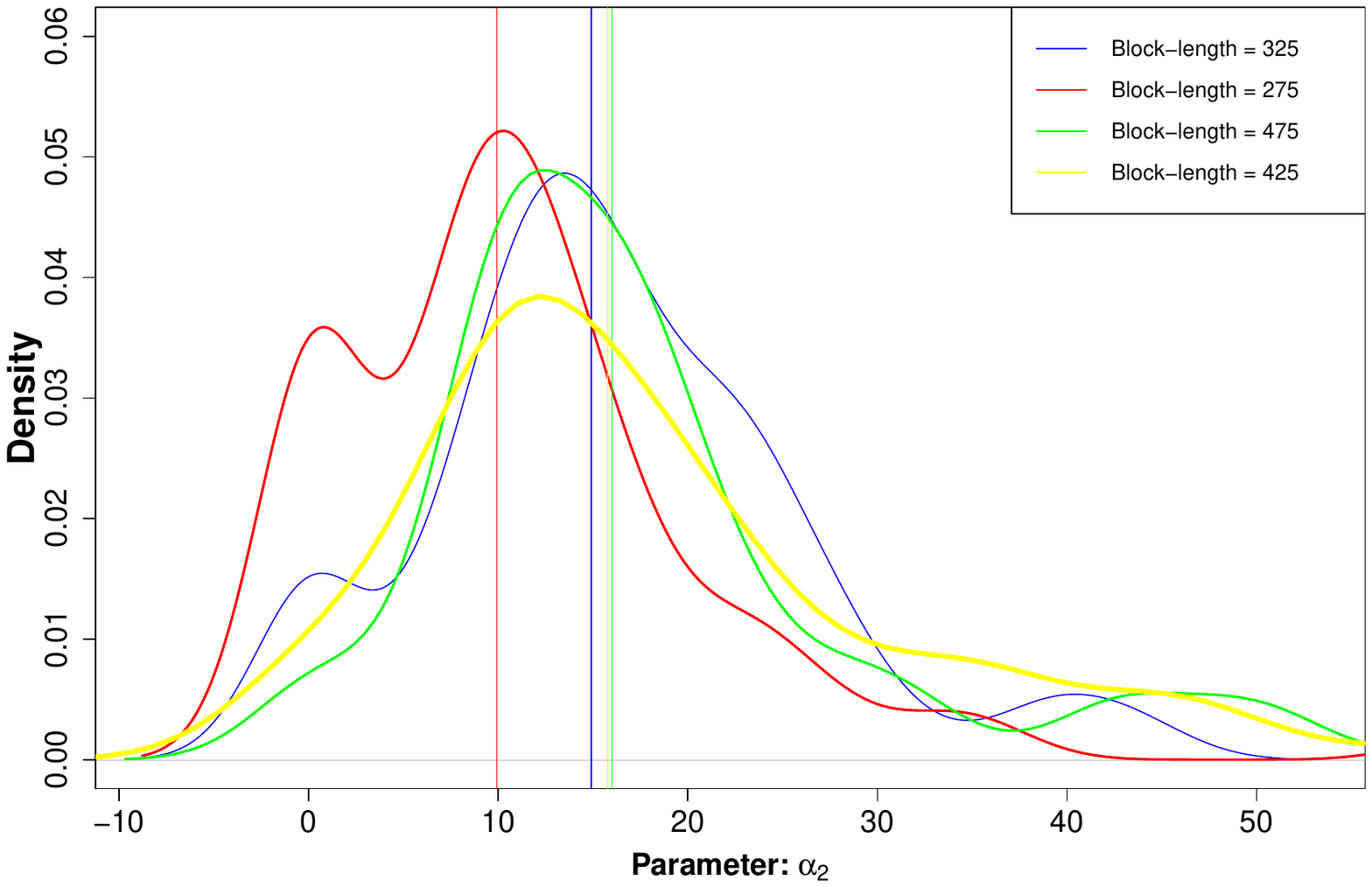}
  \caption{Distribution of parameter $\alpha_2$ by different block sizes.}
  \label{fig:saadi_distributions_block_sizes_alpha_2}
     \end{subfigure}
     \vfill
     \begin{subfigure}[b]{0.49\textwidth}
     \centering
     \includegraphics[width=0.99\textwidth]{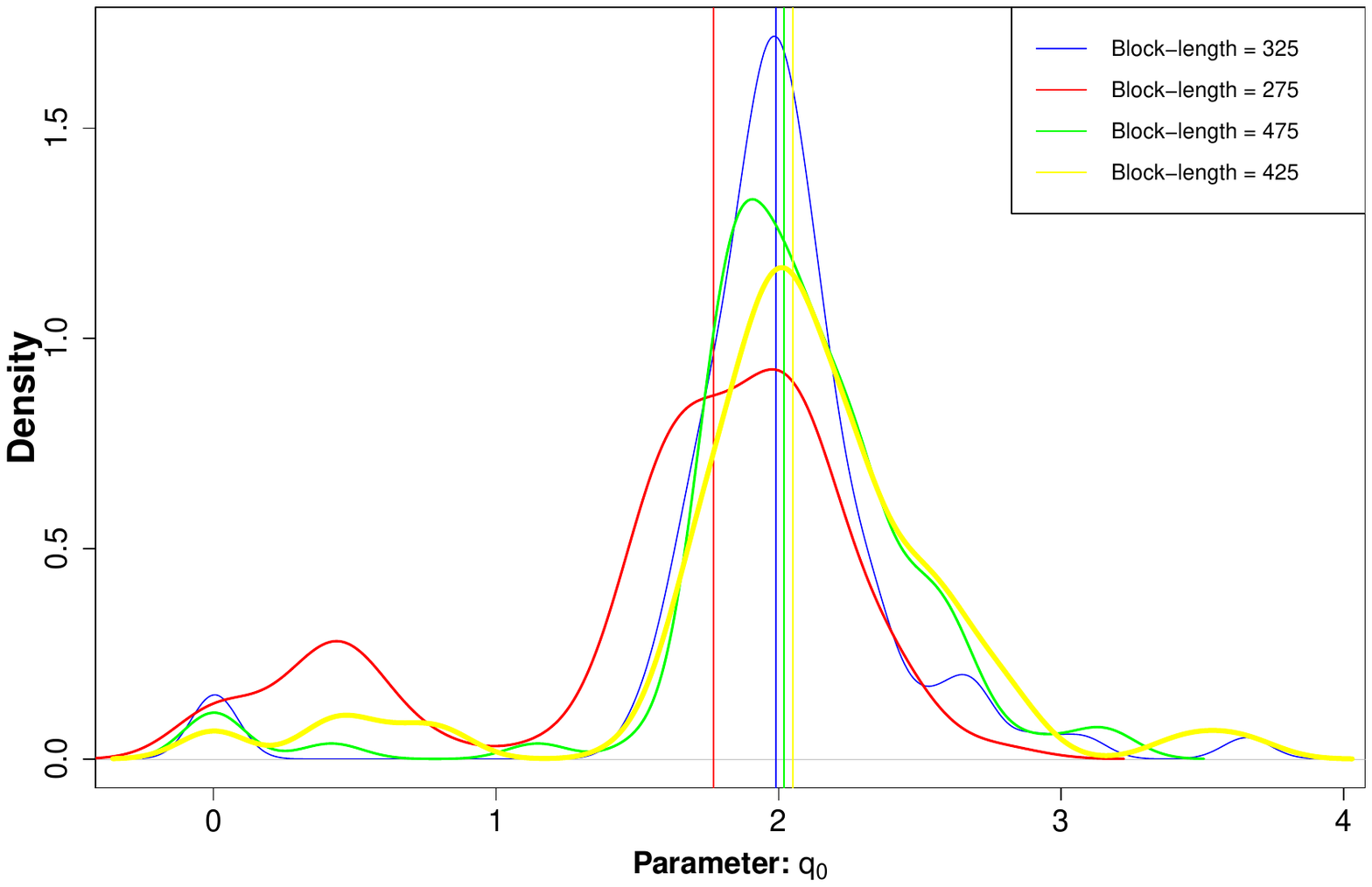}
  \caption{Distribution of parameter $q_0$ by different block sizes.}
  \label{fig:saadi_distributions_block_sizes_q_0}
     \end{subfigure}
     \hfill
     \begin{subfigure}[b]{0.49\textwidth}
     \centering
  \includegraphics[width=0.99\textwidth]{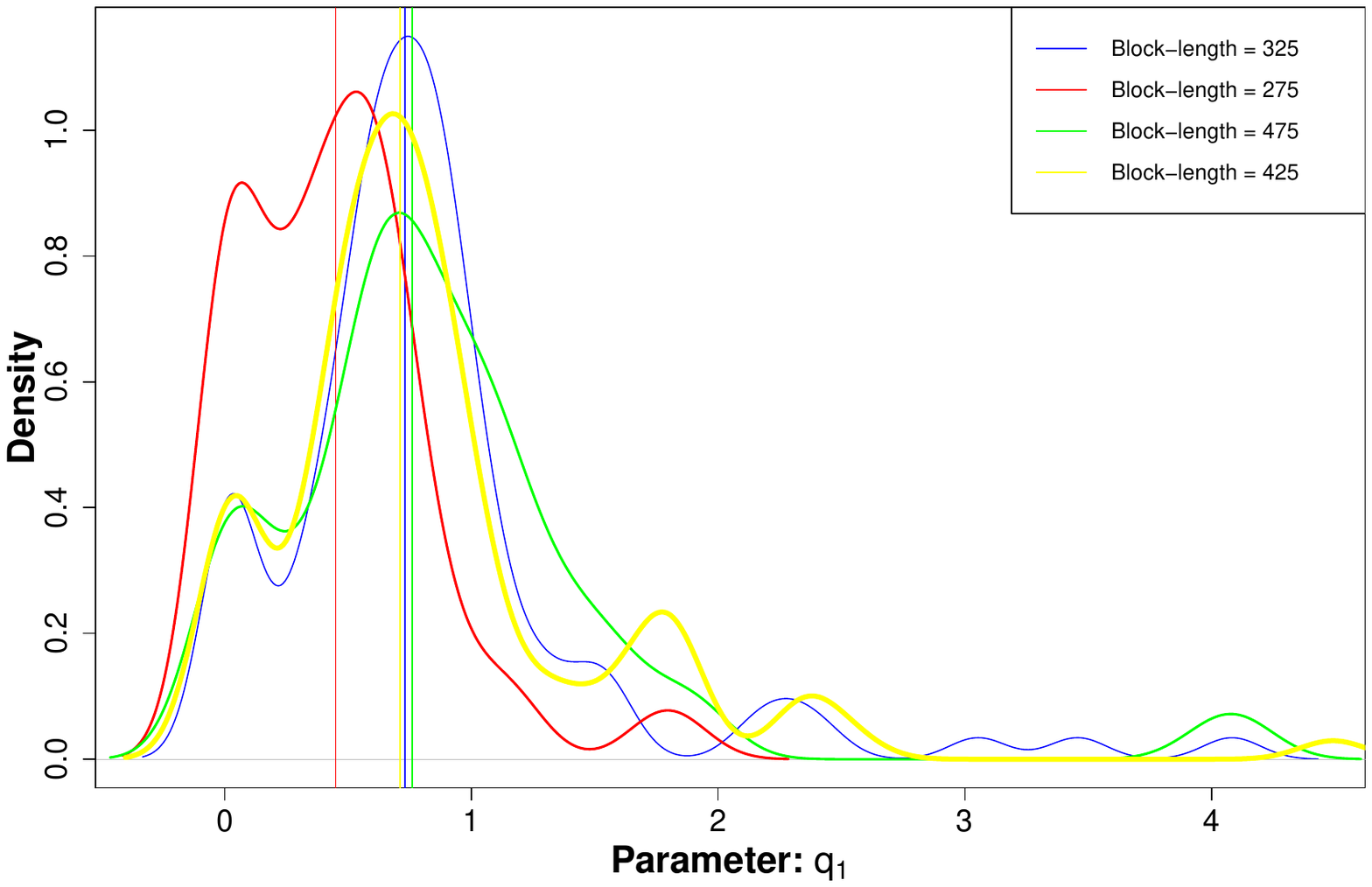}
  \caption{Distribution of parameter $q_1$ by different block sizes.}
  \label{fig:saadi_distributions_block_sizes_q_1}
     \end{subfigure}
     \caption{Distribution of parameters $\alpha_2$, $q_0$ and $q_1$ by different block sizes.}
    \label{fig:saadi_distributions_block_sizes}
\end{figure*}

\begin{figure*}[h]
\begin{minipage}{.88\textwidth}
\centering
     \begin{subfigure}[b]{0.49\textwidth}
     \centering
     \includegraphics[width=0.99\textwidth]{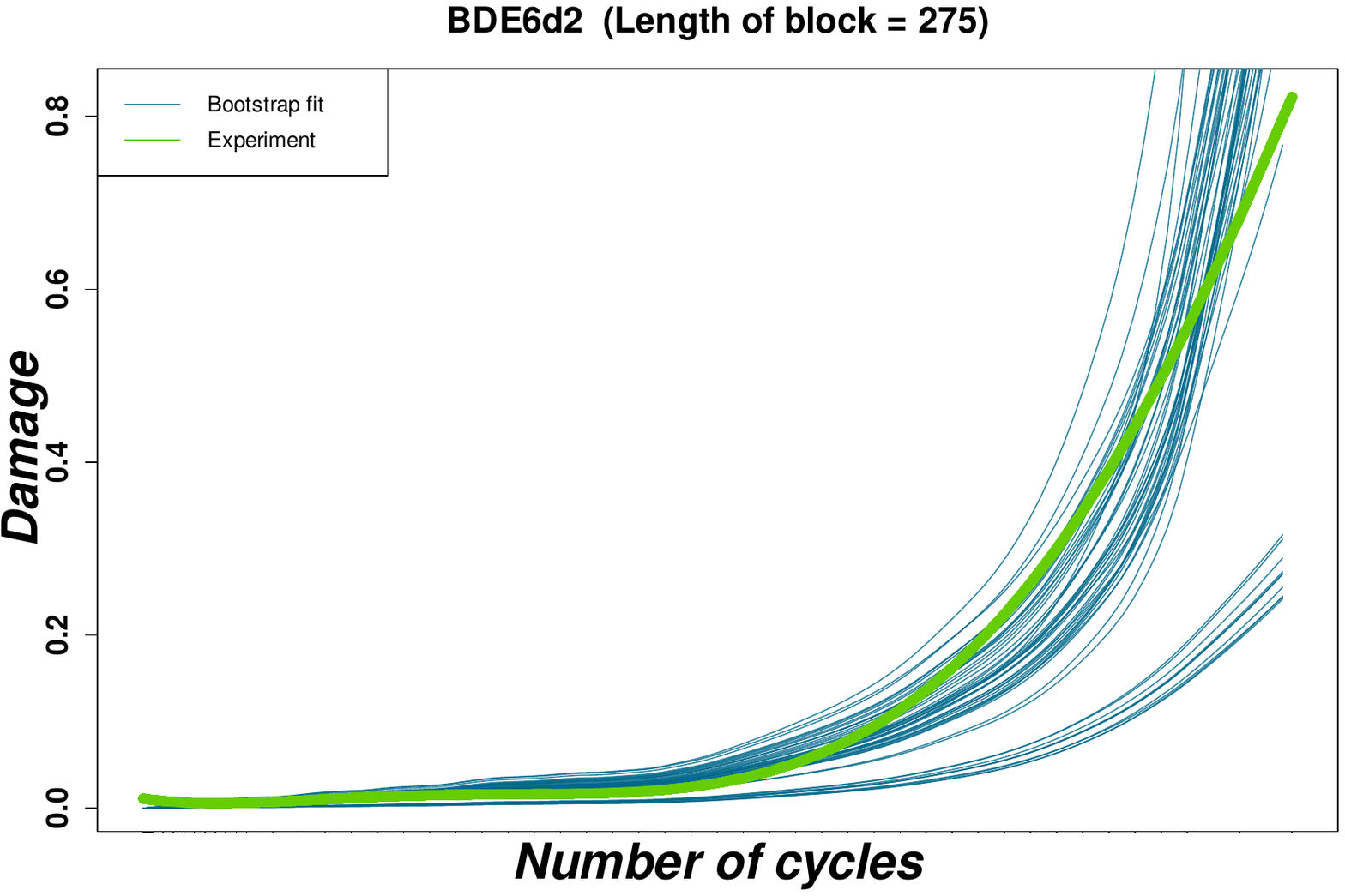}
     \end{subfigure}
          \begin{subfigure}[b]{0.49\textwidth}
     \centering
     \includegraphics[width=0.99\textwidth]{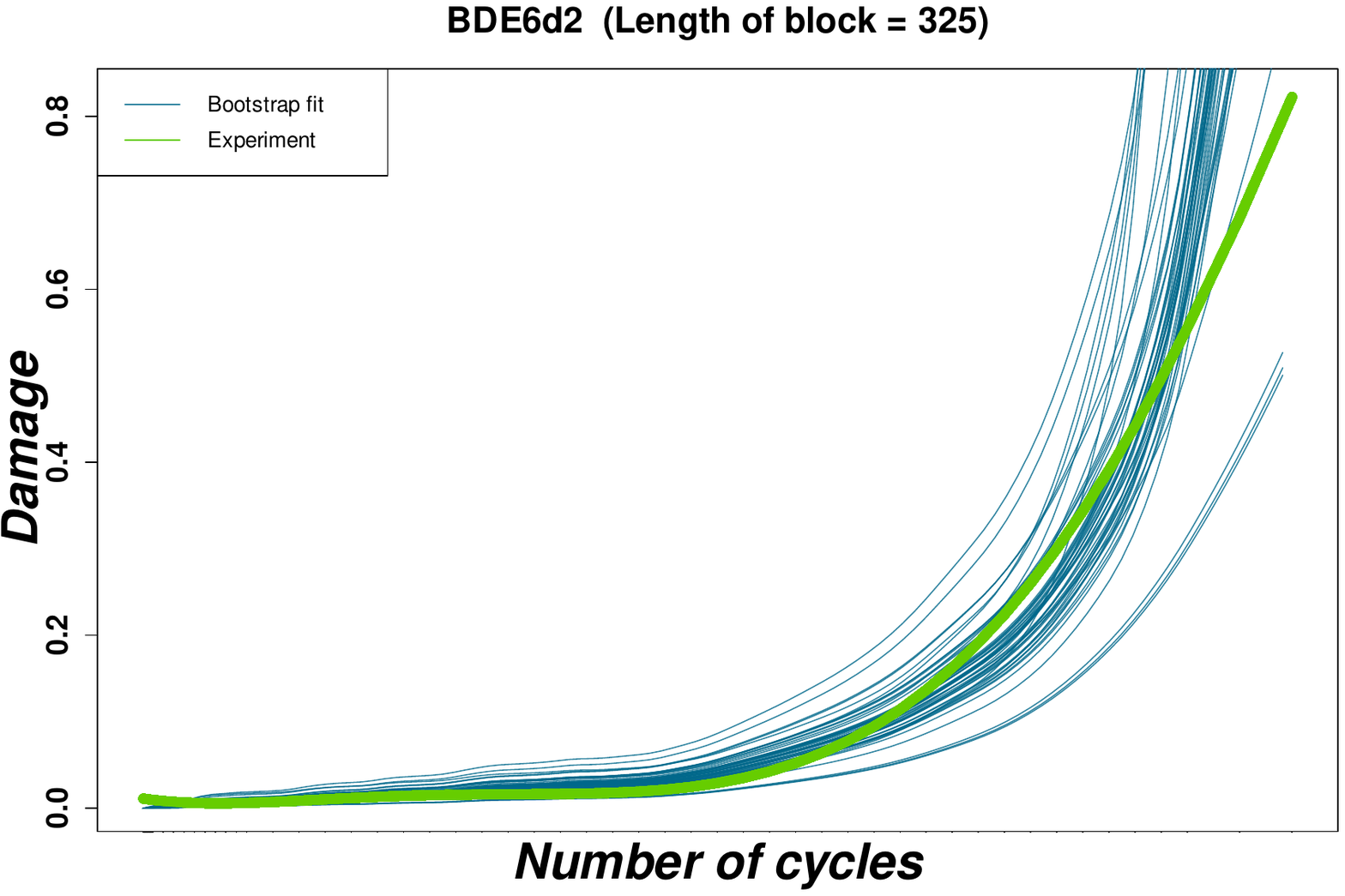}
     \end{subfigure}
     \vfill
     \begin{subfigure}[b]{0.49\textwidth}
     \centering
     \includegraphics[width=0.99\textwidth]{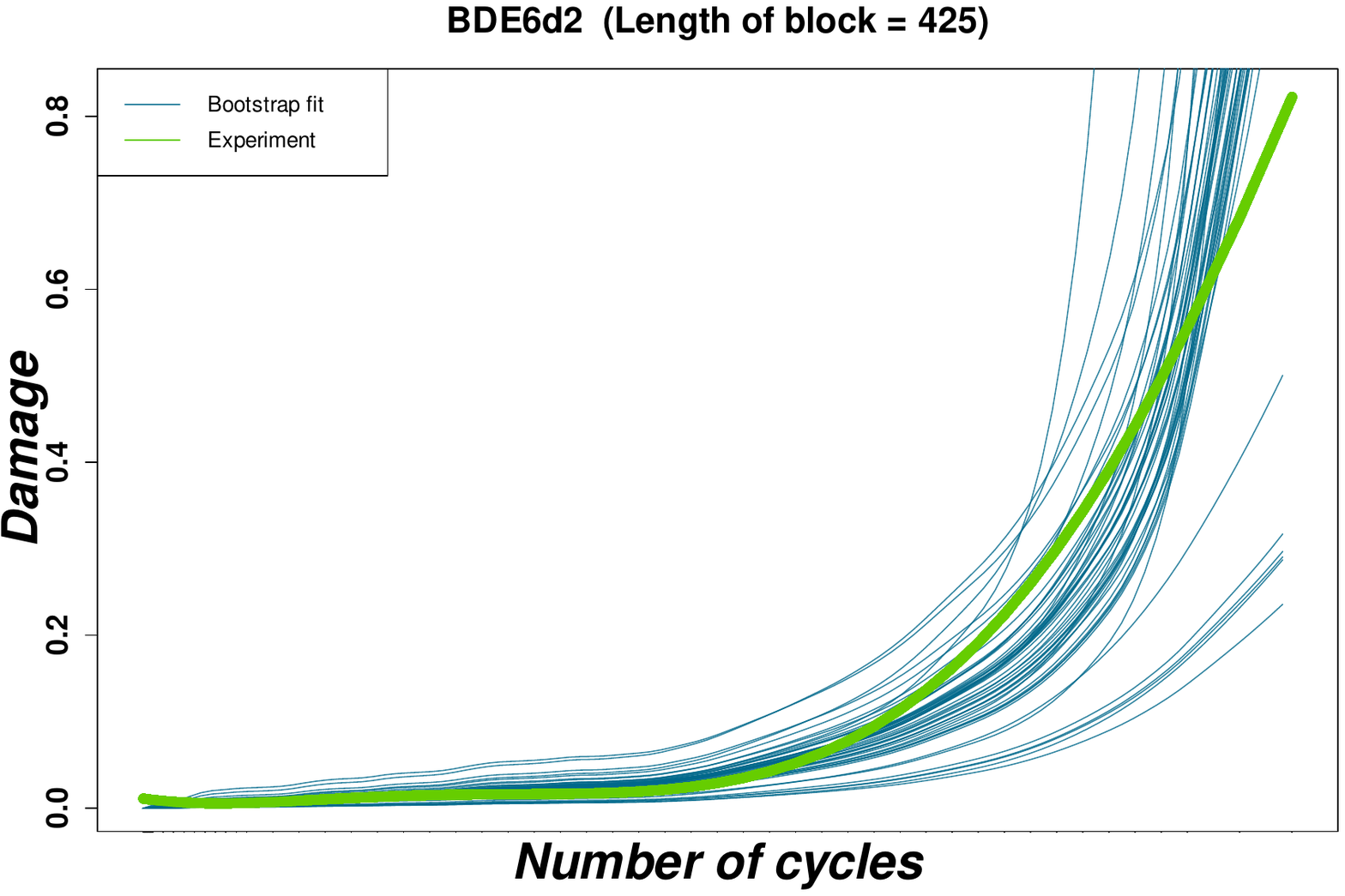}
     \end{subfigure}
     \hfill
     \begin{subfigure}[b]{0.49\textwidth}
     \centering
  \includegraphics[width=0.99\textwidth]{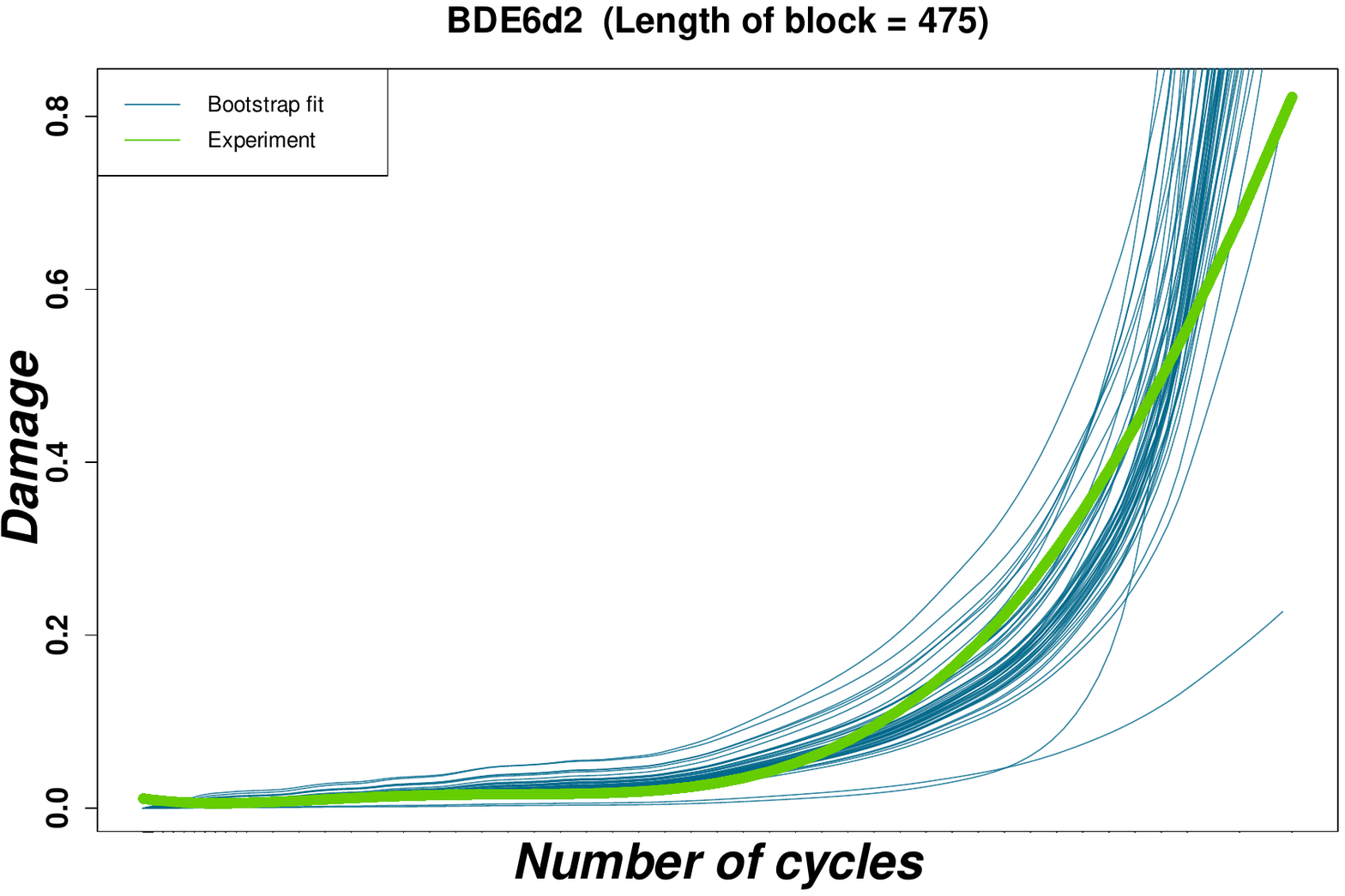}
     \end{subfigure}
     \caption{Damage curves by different block-sizes (BDE6d2).}
   \label{fig:saadi_bootstrapping_block_size_2}
   \end{minipage}
\end{figure*}

\begin{figure*}[h]
 \begin{minipage}{.88\textwidth}
\centering
     \begin{subfigure}[b]{0.49\textwidth}
     \centering
     \includegraphics[width=0.99\textwidth]{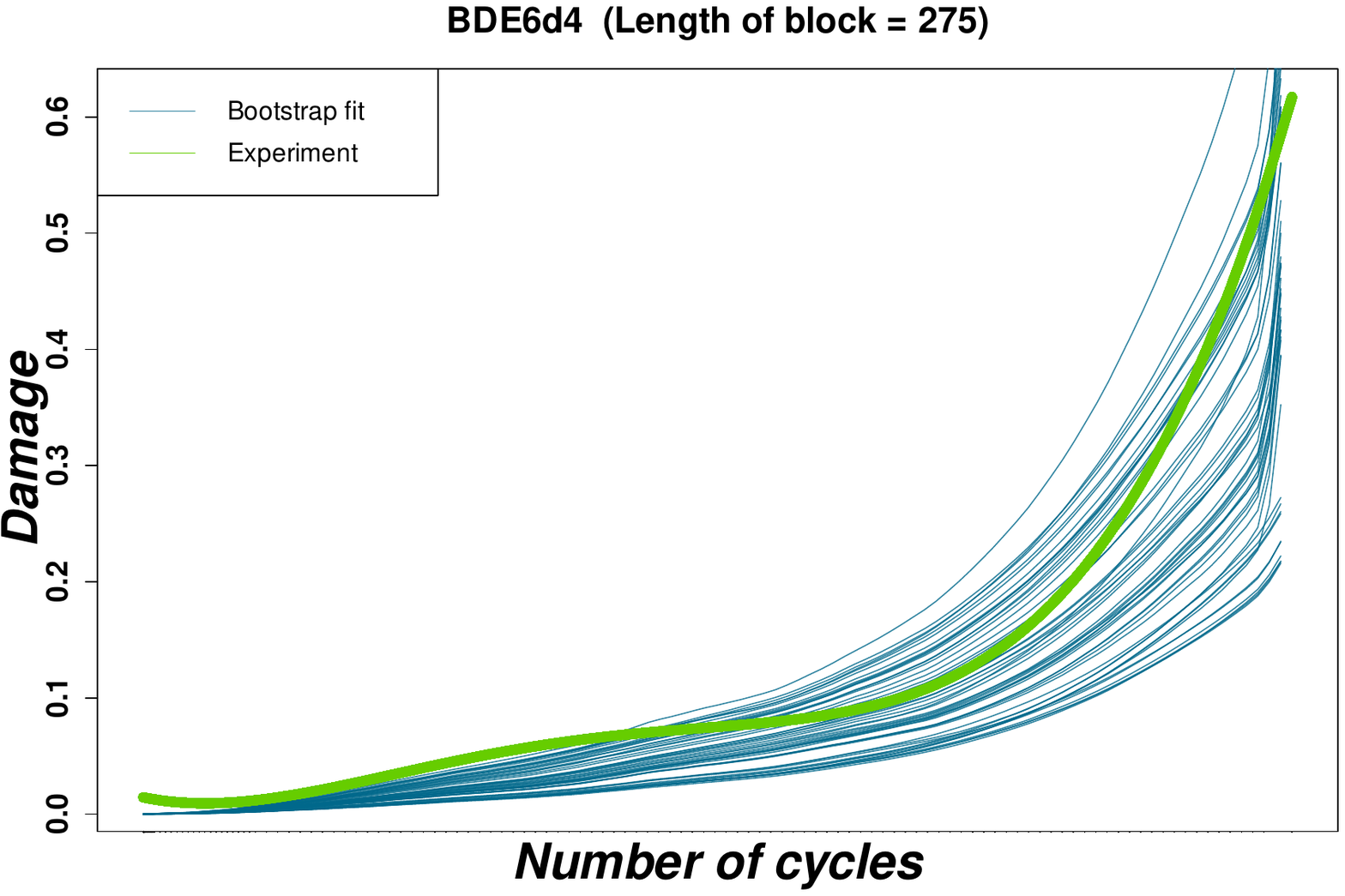}
     \end{subfigure}
          \begin{subfigure}[b]{0.49\textwidth}
     \centering
     \includegraphics[width=0.99\textwidth]{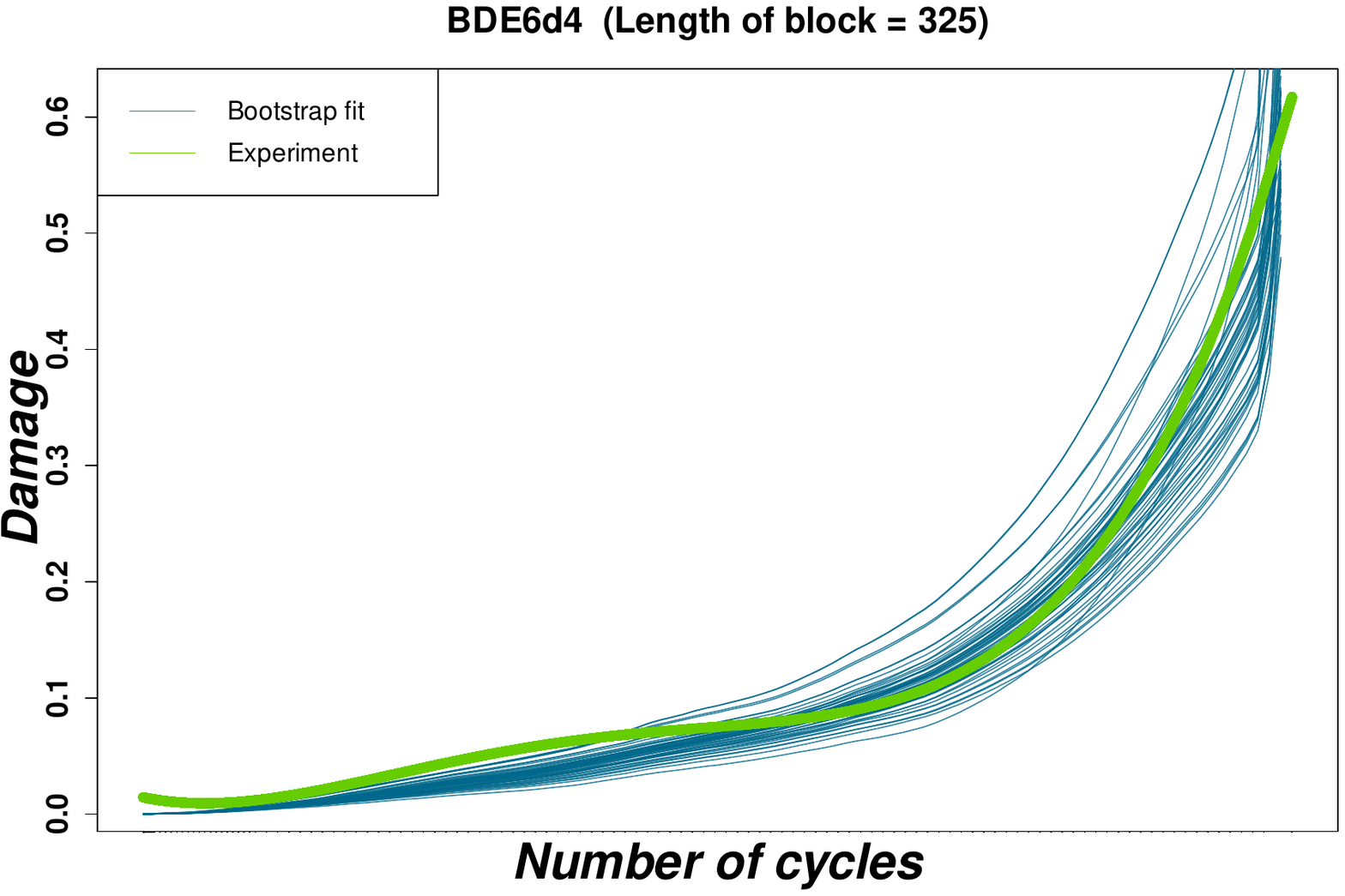}
     \end{subfigure}
     \vfill
     \begin{subfigure}[b]{0.49\textwidth}
     \centering
     \includegraphics[width=0.99\textwidth]{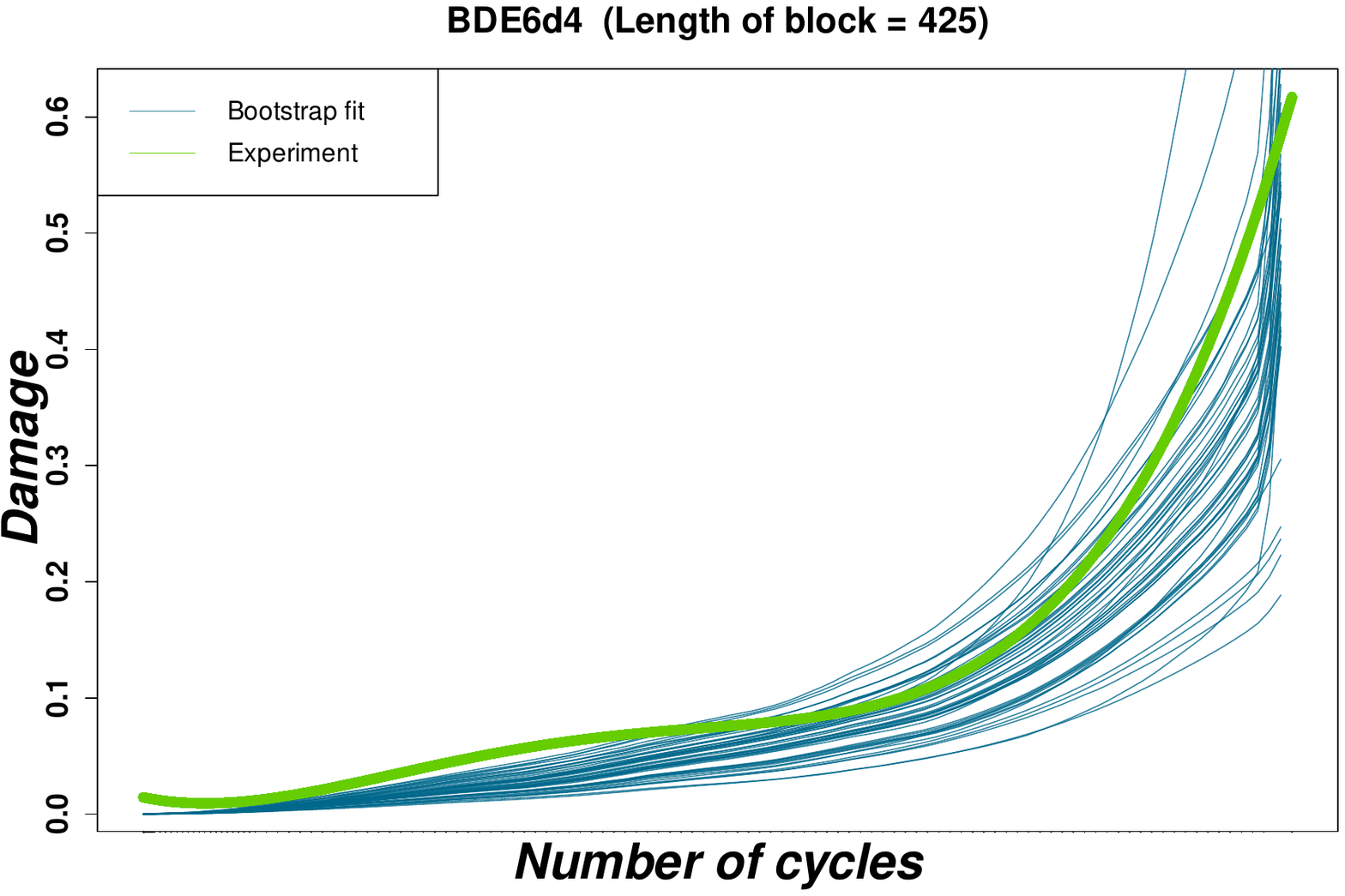}
     \end{subfigure}
     \hfill
     \begin{subfigure}[b]{0.49\textwidth}
     \centering
  \includegraphics[width=0.99\textwidth]{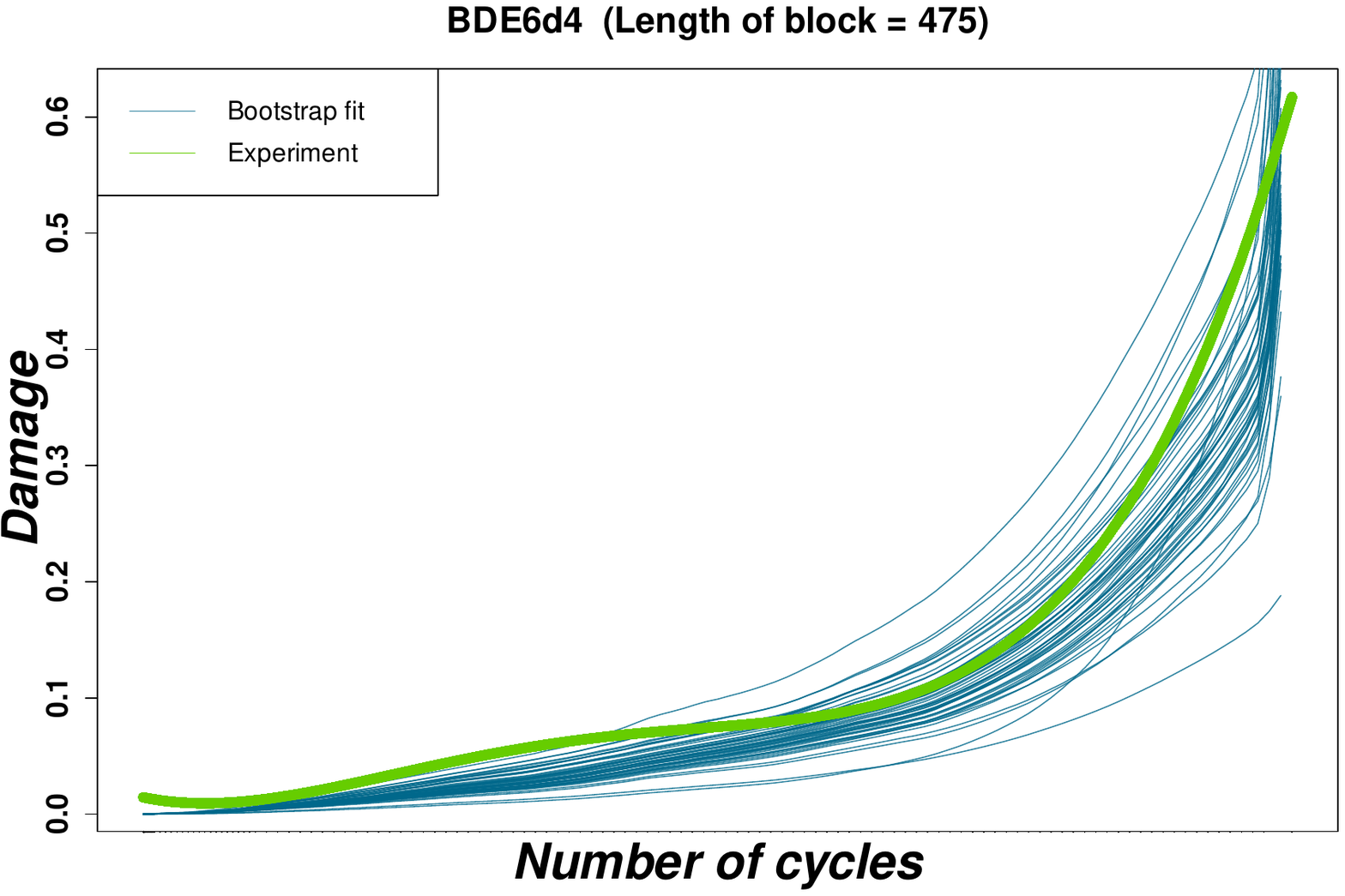}
     \end{subfigure}
     \caption{Damage curves by different block-sizes (BDE6d4).}
   \label{fig:saadi_bootstrapping_block_size_4}
    \end{minipage}
\end{figure*}

Furthermore, the bock length influence regarding the model parameters itself are 
shown in figure \ref{fig:saadi_distributions_block_sizes_alpha_2}, 
\ref{fig:saadi_distributions_block_sizes_q_0}, 
\ref{fig:saadi_distributions_block_sizes_q_1} for 100 bootstrap simulations each. 
No significant characteristic differences can be seen in the distribution of the bootstrap curves. Also, we observe significant overlap in the model parameter distributions. It can therefore be concluded that the influence of block lengths can be neglected to a first approximation.

\end{appendices}
%%%%%%%%%%%%%%%%%%%%%%%%%%%%%%%%%%%%%%%%%%%%%%%%%%%%%%%%%%%%%%%%%%%%%%%%
%  References
%%%%%%%%%%%%%%%%%%%%%%%%%%%%%%%%%%%%%%%%%%%%%%%%%%%%%%%%%%%%%%%%%%%%%%%%
\clearpage
\subsection{References}
% Please note that the files \textsf{SageH.bst} and \textsf{SageV.bst} are included with the class file
% for those authors using \BibTeX.
% The files work in a completely standard way, and you just need to uncomment one of the lines in the below example depending on what style you require:

% Harvard (name/date)
% ----> For Harvard Style use SageH
\bibliographystyle{SageH}
%
%Vancouver (numbered)
% https://journals.sagepub.com/author-instructions/IJD#ReferenceStyle
% International Journal of Damage Mechanics adheres to the Sage Vancouver reference style.
% --> 
% \bibliographystyle{SageV}
\bibliography{bibfile}
% and remember to add the relevant option 
% to the \verb+\documentclass[]{sagej}+ line as listed in Table~\ref{T1}. 
\end{document}